\documentclass[aps,superscriptaddress,preprintnumbers,showpacs]{revtex4}
\usepackage{amssymb}
\usepackage{dcolumn}
\usepackage{bm}
\usepackage{graphicx}
\usepackage{booktabs}
\usepackage{multirow}
\usepackage{color}

\begin{document}

\newcommand*{\PKU}{School of Physics and State Key Laboratory of Nuclear Physics and
Technology, Peking University, Beijing 100871,
China}\affiliation{\PKU}
\newcommand*{\CICQM}{Collaborative Innovation Center of Quantum Matter, Beijing, China}\affiliation{\CICQM}
\newcommand*{\CHEP}{Center for High Energy Physics, Peking University, Beijing 100871, China}\affiliation{\CHEP}

\title{The Majorana neutrino mass matrix indicated by the current data}

\author{Xinyi Zhang}\affiliation{\PKU}
\author{Bo-Qiang Ma}\email{mabq@pku.edu.cn}\affiliation{\PKU}\affiliation{\CICQM}\affiliation{\CHEP}

\begin{abstract}
The Majorana neutrino mass matrix combines information from the neutrino masses and the leptonic mixing in the flavor basis. Its invariance under some transformation matrices indicates the existence of certain residual symmetry. We offer an intuitive display of the structure of the Majorana neutrino mass matrix, using the whole set of the oscillation data. The structure is revealed depending on the lightest neutrino mass. We find that there are three regions with distinct characteristics of structure. A group effect and the $\mu$-$\tau$ exchange symmetry are observed. Six types of texture non-zeros are shown. Implications for flavor models are discussed.

\end{abstract}


\maketitle

\section{Introduction}
\label{sec:intro}
Neutrino oscillation has been well established by various experiments regarding solar~\cite{solar1,solar2}, atmospheric~\cite{atmospheric1,atmospheric2,atmospheric3,atmospheric4}, and reactor~\cite{double chooz,dayabay,reno,kamland} neutrinos. The oscillation is caused by the mismatch of the neutrino mass eigenstates with the flavor eigenstates.

Generally, the oscillation probability measured by the oscillation experiments depends on the neutrino energy, the distance of flight, the mixing matrix elements, and the squared mass differences. The physical parameters in the oscillation probability are three mixing angles $\theta_{12},\theta_{23},\theta_{13}$ and a CP phase $\delta$ coming from the
Pontecorvo-Maki-Nakawaga-Sakata (PMNS) matrix~\cite{p,mns} and two squared mass differences $\Delta m_{21}^2,\Delta m_{31(2)}^2$ show up in two observed frequencies of the oscillation probability, where $\Delta m_{ij}^2\equiv m_i^2-m_j^2$ and $m_i$ denotes the $i$th mass eigenstate of the neutrino. These parameters involve information on both mass and mixing. For Majorana neutrinos, that is, when the neutrinos and their antiparticle partners are identical to each other, the Lagrangian part regarding the leptonic masses and mixing is
\begin{eqnarray}
\mathcal{L}=\frac{g}{\sqrt{2}}\bar l_L\gamma^\mu\nu_LW_\mu^- +\bar l_L  m_l E_R+\frac{1}{2}\bar \nu_L \rm M \nu_L^c+H.c.,
\end{eqnarray}
where $l_L=(e_L,\mu_L,\tau_L)^T$ denotes the left-handed charged leptons, $\nu_L=(\nu_{eL},\nu_{\mu L},\nu_{\tau L})^T$ denotes the left-handed neutrinos, $E_R=(e_R,\mu_R,\tau_R)^T$ denotes the right-handed charged leptons, and $\rm M$ is the Majorana mass matrix. We choose the flavor basis, that is, the charged lepton mass eigenstates coincide with the flavor eigenstates, then the information about mixing is contained solely in the Majorana mass matrix, together with the neutrino mass information, which is relevant to the constituents of the universe~\cite{SB,Hwang:2012vx}. Whether neutrinos are Majorana or Dirac particles is also inspiring for searching some phenomenological symmetries in condensed matter and optical physics~\cite{SC}.

From a theoretical viewpoint, a Majorana neutrino mass matrix stemming from a dimensional-$5$ Weinberg operator~\cite{Weinberg:1979sa}, that is, $\frac{a_{ij}}{\Lambda}l_ihl_jh$ where $l$ denotes a lepton doublet and $h$ denotes the Higgs doublet, can give us a natural explanation of the smallness of the neutrino mass~\cite{xing}, as explored by various seesaw models~\cite{seesaw1,seesaw2,seesaw3,seesaw4,seesaw5,seesaw6} where the lightness of neutrino masses is naturally accounted for after integrating the heavy messenger fields. Besides, the invariance of the Majorana mass matrix under some transformation matrices, that is, $G^T \mathrm{M} G=\rm M$ (where $G$ denotes a group generator), shows evidences for the existence of certain flavor symmetry, and extensive works are dedicated on this issue (see~\cite{Altarelli:2010gt} for a review).

Since the oscillation experiments are continuously making efforts to better determine the oscillation parameters, we find that it is necessary to investigate the Majorana neutrino mass matrix, especially after the non-zero and relatively large value of  $\theta_{13}$~\cite{dayabay,reno}. As will be shown later, this investigation gives us the dominant structures at given $m_{\rm min}$ and enables us to make quick comparisons with the model results. Besides, one can also observe possible texture zeros directly. Regarding an investigation of the neutrino mass matrix after the measurement of $\theta_{13}$, ref.~\cite{Grimus:2012ii} uses inequalities to give the allowed range and the correlations for the absolute values of the neutrino mass matrix elements, and ref.~\cite{Bertuzzo:2013ew} investigates the correlations of the neutrino mass matrix entries by constructing probability distribution functions for each mixing parameter. This paper differs from the other two both in methodology and in emphases.

It is our primary intention to check the constraints on the structure of the Majorana mass matrix given by the current data. Using the simple relation that correlates the mass and the mixing matrix, that is, $\mathrm{M}=\rm U^\star \rm diag(m_1,m_2,m_3) \rm U^\dagger$, we reconstruct the Majorana mass matrix up to an unknown mass. This procedure is taken analytically and the errors of each parameter are carefully passed to the mass matrix entries. We exhibit all the $|\rm M_{\alpha\beta}|$ in a same plot to make an easy observation and a quick comparison.

This paper is organized as follows. First, we reconstruct the Majorana neutrino mass matrix in Section~\ref{sec:reconstruct}; then, we discuss the major features of the results and other related issues in Section~\ref{sec:discuss}; and final conclusions and discussions on the implications for the flavor models are given in Section~\ref{sec:conclusion}.

\section{Reconstruct the Majorana neutrino mass matrix}
\label{sec:reconstruct}
\subsection{Input}
The Majorana neutrino mass matrix can be obtained through
\begin{eqnarray}
\rm M=\rm U^* \rm Diag\{m_1,m_2,m_3\} \rm U^\dagger.
\end{eqnarray}
In the flavor basis, where the charged lepton mass matrix is diagonal, the matrix $\rm U$ diagonalizing the mass matrix $\rm M$ is the PMNS matrix. Since
\begin{eqnarray}
\rm U=\rm V \rm P,\quad \rm P=Diag\{1,e^{i\frac{\alpha_{21}}{2}},e^{i\frac{\alpha_{31}}{2}}\},
\end{eqnarray}
 and adopting a standard parametrization~\cite{CK} for $\rm V$,

\begin{eqnarray}
\rm V=\left(
\begin{array}{ccc}
c_{12}c_{13} & s_{12}c_{13} & s_{13}e^{-i\delta}         \\
-s_{12}c_{23}-c_{12}s_{23}s_{13}e^{i\delta} & c_{12}c_{23}-s_{12}s_{23}s_{13}e^{i\delta} & s_{23}c_{13} \\
s_{12}s_{23}-c_{12}c_{23}s_{13}e^{i\delta} & -c_{12}s_{23}-s_{12}c_{23}s_{13}e^{i\delta} & c_{23}c_{13}\\
\end{array}
\right),
\end{eqnarray}
we get the nine parameters in need to determine the Majorana mass matrix:
\begin{eqnarray*}
\theta_{12},\theta_{23},\theta_{13},\delta,\alpha_{21},\alpha_{31},m_1,m_2,m_3.
\end{eqnarray*}

The oscillation experiments measure six of the nine parameters, which are three mixing angles, the Dirac CP-violating phase $\delta$, and two squared mass differences. Although the Dirac CP-violating phase is not determined by the experiments by now, the global fit results give us some clues on its value (see Tables~\ref{tab:i}~and~\ref{tab:ii}, also~\cite{valle}). The recent T2K results~\cite{T2K} also suggest a maximal CP-violation with a minus sign, that is, $\delta=-90^\circ~(270^\circ)$, which is consistent with the global fit at the confidence level of $1$ \emph{$\sigma$}.

The Majorana phases contained in the matrix $\rm P$ do not manifest them in the oscillation experiments. Furthermore, we have no knowledge of them. To do the calculation, the two Majorana phases are set to zero by hand. Some comments about this treatment will be shown in the following section.

Using the whole set of the global fit results~\cite{glb_fit}, together with an assumption of $\alpha_{21}=\alpha_{31}=0$, we can determine the neutrino mass matrix with an unknown mass standing for the neutrino mass scale. We list the input in Table~\ref{tab:i}.   Notice that the octant of $\theta_{23}$ is not clear, and we use the first octant one just for simplicity. The explicit form of the expressions for $\rm M_{\alpha\beta}$ can be found in Section~\ref{sec:elements} of the Appendix. We also use the global fit results~\cite{Fogli:2012ua,fogli2013} as an input and list the results in Section~\ref{sec:fogli} of the Appendix.

For basis other than the flavor basis, the charged lepton mass matrix can be non-diagonal, and it can be diagonalized through a  biunitary transformation by a matrix $\rm V_L^l$, namely,
\begin{eqnarray}
\rm V_L^{l\dagger} \rm M_l \rm M_l^\dagger \rm V_L^l=Diag\{m_e^2,m_\mu^2,m_\tau^2\}.
\end{eqnarray}
In the mean time, the neutrino mass matrix can be diagonalized by
\begin{eqnarray}
\rm V_L^{\nu T} \rm M_\nu \rm V_L^\nu=Diag\{m_1,m_2,m_3\}\label{eqn:nond}
\end{eqnarray}
for Majorana neutrinos. Thus, the PMNS matrix for lepton mixing is $U_{\rm PMNS}=\rm V_L^{l \dagger} \rm V_L^\nu$. To reconstruct the Majorana neutrino mass matrix in this case, we need to specify either the charged lepton mass matrix or the $\rm V_L^l$ matrix first, then we can translate the information to $\rm V_L^\nu$ and use Eq.~\ref{eqn:nond}.

\begin{table}[tbp]
\centering
\begin{tabular}{|l|rc|}
\hline
& Best fit & $3 \sigma$ range\\ \hline
$\sin^2 \theta_{12}$ & 0.306 & 0.271 $-$ 0.346\\
$\sin^2 \theta_{23}$ & 0.446 & 0.366 $-$ 0.663\\
$\sin^2 \theta_{13}$ & 0.0231& 0.0173 $-$ 0.0288\\
$\Delta m_{21}^2 (10^{-5}\rm{eV}^2)$ & 7.45 & 6.98 $-$ 8.05\\
$\Delta m_{31}^2 (10^{-3}\rm{eV}^2)$ & 2.417& 2.247 $-$ 2.623\\
$\Delta m_{32}^2 (10^{-3}\rm{eV}^2)$ & $-$2.411& $-$2.602 $-$ $-$2.226\\
$\delta(\rm{degree})$ & 266 & 1 $-$ 360\\
\hline
\end{tabular}
\caption{\label{tab:i} Global fit results from ref.~\cite{glb_fit}.}
\end{table}

\begin{table}[tbp]
\centering
\begin{tabular}{|l|rc|}
\hline
& Best fit & $3 \sigma$ range\\ \hline
$\sin^2 \theta_{12}~(\rm{N}~or~\rm{I})$ & 0.308 & 0.259 $-$ 0.359\\
$\sin^2 \theta_{23}~(\rm{N})$ & 0.437 & 0.374 $-$ 0.626\\
$\sin^2 \theta_{23}~(\rm{I})$ & 0.455 & 0.380 $-$ 0.641\\
$\sin^2 \theta_{13}~(\rm{N})$ & 0.0234&0.0176 $-$ 0.0259\\
$\sin^2 \theta_{13}~(\rm{I})$ & 0.0240&0.0178 $-$ 0.0298\\
$\delta m^2 (10^{-5}\rm{eV}^2)$ & 7.54 & 6.99 $-$ 8.18\\
$\Delta m^2 (10^{-3}\rm{eV}^2)~(\rm{N})$ & 2.43 & 2.23 $-$ 2.61\\
$\Delta m^2 (10^{-3}\rm{eV}^2)~(\rm{I})$ & $-$2.38 & $-$2.56 $-$ $-$2.19\\
$\delta~(\rm{N})$ & 1.39$\pi$ & 0 $-$ 2$\pi$\\
$\delta~(\rm{I})$ & 1.31$\pi$ & 0 $-$ 2$\pi$\\
\hline
\end{tabular}
\caption{\label{tab:ii} Global fit results from ref.~\cite{fogli2013}. \rm N (\rm I) stands for the normal (inverted) ordering.}
\end{table}

\subsection{Extreme case}
If the neutrino masses are highly hierarchical, that is, the lightest neutrino mass is small enough to be neglected, the other two masses can be determined by the two squared mass differences. We consider that a percent level quantity is small enough and make a rough estimate for a to-be-neglected mass scale. Since the small squared mass difference is of $\mathcal{O}(10^{-5})\rm~eV^2$, the upper limit for the mass-to-be-neglected is the squared root of $10^{-7}\rm~eV^2$, which is of $\mathcal{O}(10^{-4})\rm~eV$.

In the normal hierarchical case, that is, $m_1\leq 10^{-4}~\rm{eV}$, we omit it and get
\begin{eqnarray}
m_2&\simeq&\sqrt{\Delta m_{21}^2}=0.0086~ \rm{eV};\\ \nonumber
m_3&\simeq&\sqrt{\Delta m_{31}^2}=0.0492~ \rm{eV};\\ \nonumber
\rm |M|&\simeq&\left(
\begin{array}{ccc}
0.0015 & 0.0058 & 0.0065\\
.      & 0.0248 & 0.0209\\
.      & .      & 0.0292\\
\end{array}
\right),
\end{eqnarray}
where ``$.$'' stands for the symmetric counterparts of $|\rm M_{\alpha\beta}|$.

In the inverted hierarchical case, where $m_3\leq 10^{-4}~\rm{eV}$ and can be neglected, we have
\begin{eqnarray}
m_1&\simeq& m_2 \simeq \sqrt{\Delta m_{32}^2}=0.0491~ \rm{eV};\\ \nonumber
\rm |M|&\simeq&\left(
\begin{array}{ccc}
0.0480 & 0.0049 & 0.0055\\
.      & 0.0267 & 0.0250\\
.      & .      & 0.0213\\
\end{array}
\right).
\end{eqnarray}

Similarly, if the neutrinos are massive enough to neglect the mass differences, they can be quasi-degenerate. In this case,
\begin{eqnarray}
m_1 &\simeq& m_2 \simeq m_3\simeq 0.1~ \rm{eV};\\ \nonumber
\rm |M|&\simeq&\left(
\begin{array}{ccc}
0.0955 & 0.0200 & 0.0221\\
.      & 0.0980 & 0.0023\\
.      & .      & 0.0975\\
\end{array}
\right).
\end{eqnarray}

\subsection{General case}
\begin{figure*}
\begin{minipage}{\textwidth}
   \centering
\includegraphics[width=.45\textwidth]{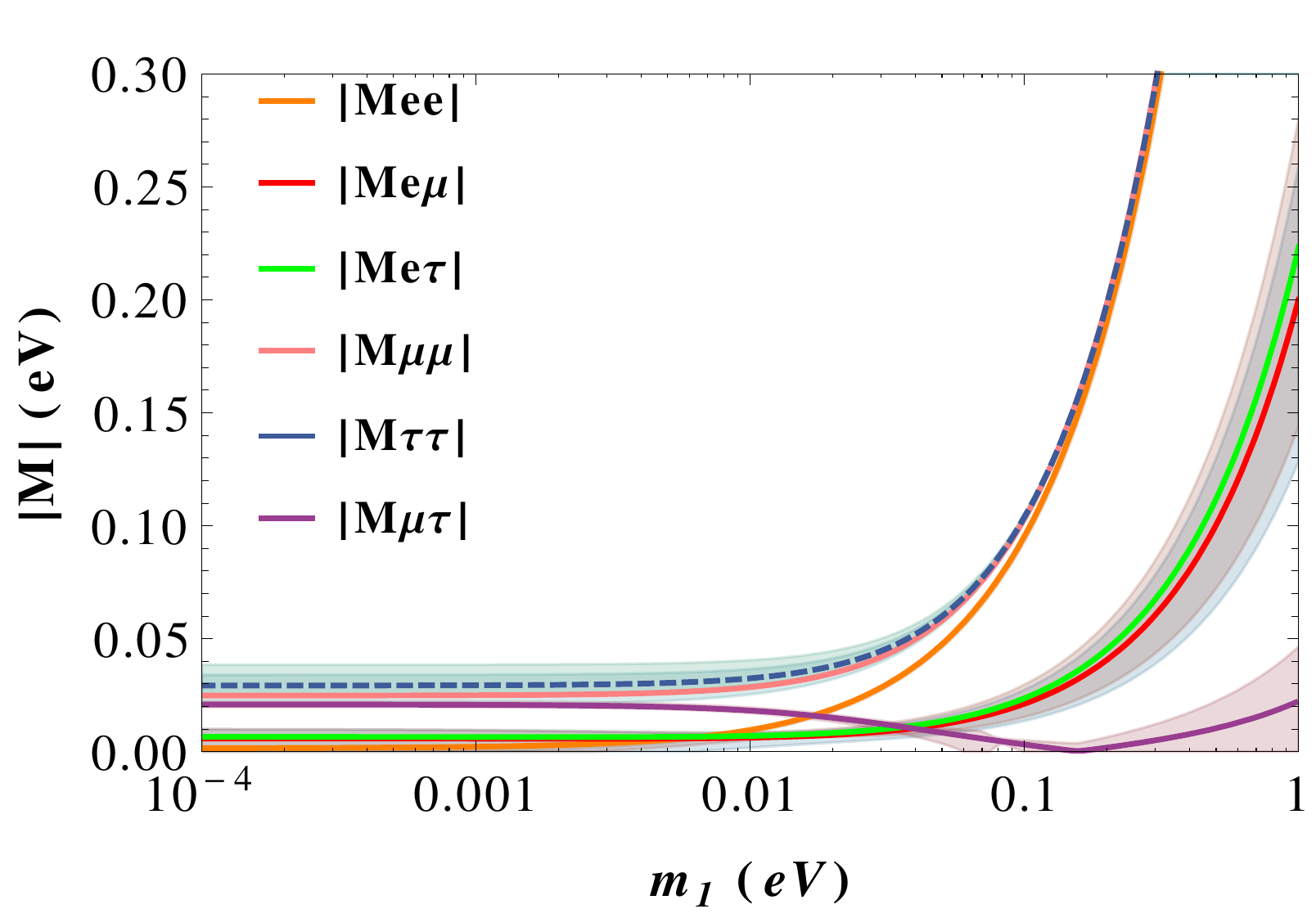}
\includegraphics[width=.45\textwidth]{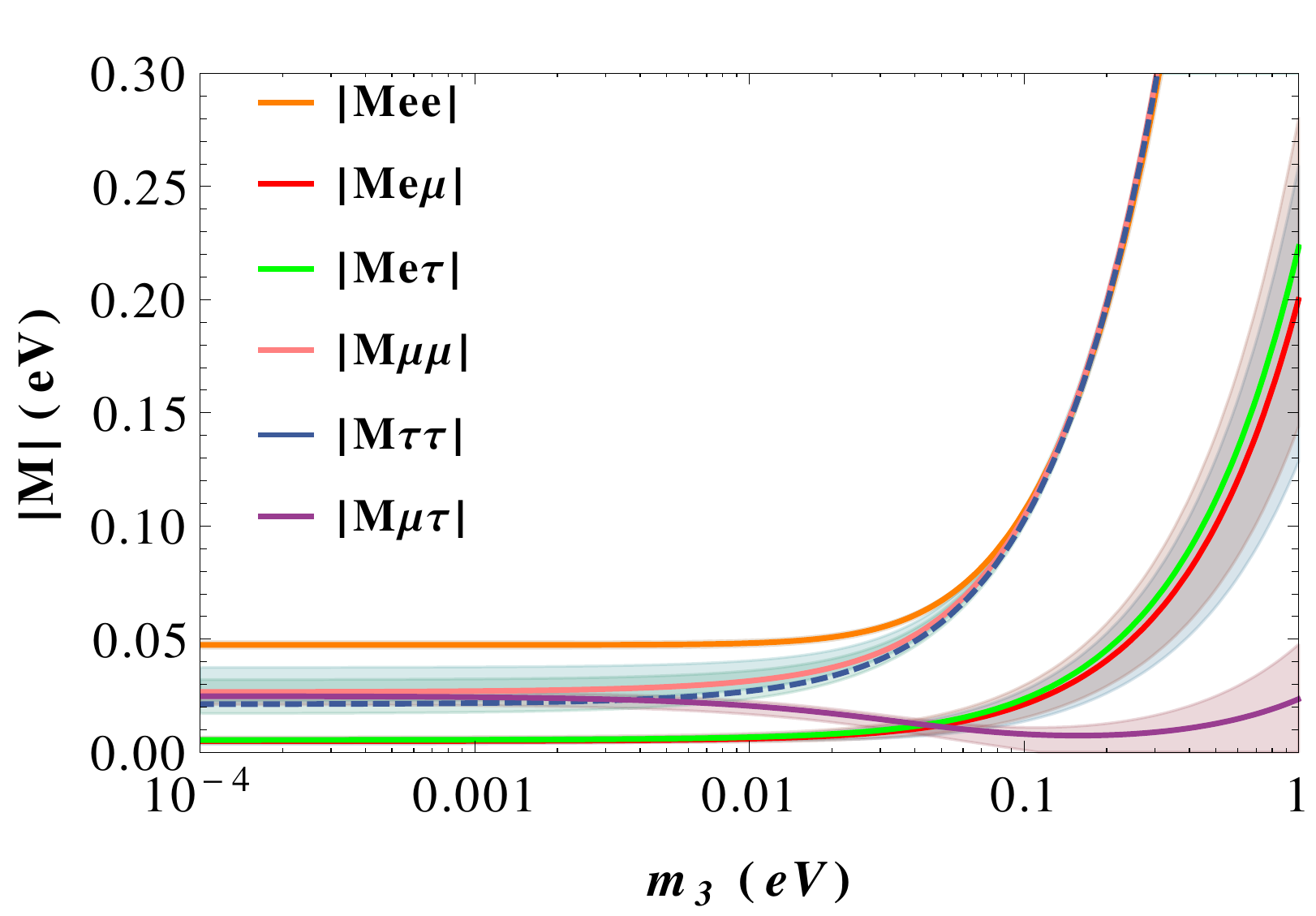}
           \caption{\label{fig:i} $|\rm M_{\alpha\beta}|$ as a function of the lightest neutrino mass. The left one corresponds to the normal ordering, and the right one corresponds to the inverted ordering. The bands are calculated with the $3\sigma$ error ranges of the inputting parameters.}
\end{minipage}
\end{figure*}

We move on to a general case where the unknown lightest neutrino mass is taken to be a variable. We plot the dependence of the $|\rm M_{\alpha\beta}|$ on $m_{\rm min}$ in Figure~\ref{fig:i}.

According to the stability of the relative magnitudes of different $|\rm M_{\alpha\beta}|$, we recognize three regions with distinct characteristics.

\begin{enumerate}
\item For a $m_{\rm min} <0.002~\rm{eV}$, the relative magnitudes of $|\rm M_{\alpha\beta}|$ are stable. We label it as Region \uppercase\expandafter{\romannumeral1}.\\
\item For $0.002~\rm{eV} <\emph{m}_{\rm min} <0.1~\rm{eV}$, the relative magnitudes change several times (there are several crossings), and the crossing points differ in the normal ordering and the inverted ordering. This region is called Region \uppercase\expandafter{\romannumeral2}.\\
\item For $m_{\rm min} >0.1~\rm{eV}$, the relative magnitudes of $|\rm M_{\alpha\beta}|$ are stable again. It is labeled as Region \uppercase\expandafter{\romannumeral3}.\\
\end{enumerate}

As can be seen from Figure~\ref{fig:i}, this regional division works for both orderings. In Region \uppercase\expandafter{\romannumeral1} and Region \uppercase\expandafter{\romannumeral3}, the stability of the relative magnitudes of $|\rm M_{\alpha\beta}|$ indicates the existence of one dominant structure. The large error ranges in Region \uppercase\expandafter{\romannumeral3} are mainly due to the complex dependence with various inputting parameters.

\begin{figure*}
\begin{minipage}{\textwidth}
   \centering
   \includegraphics[width=.45\textwidth]{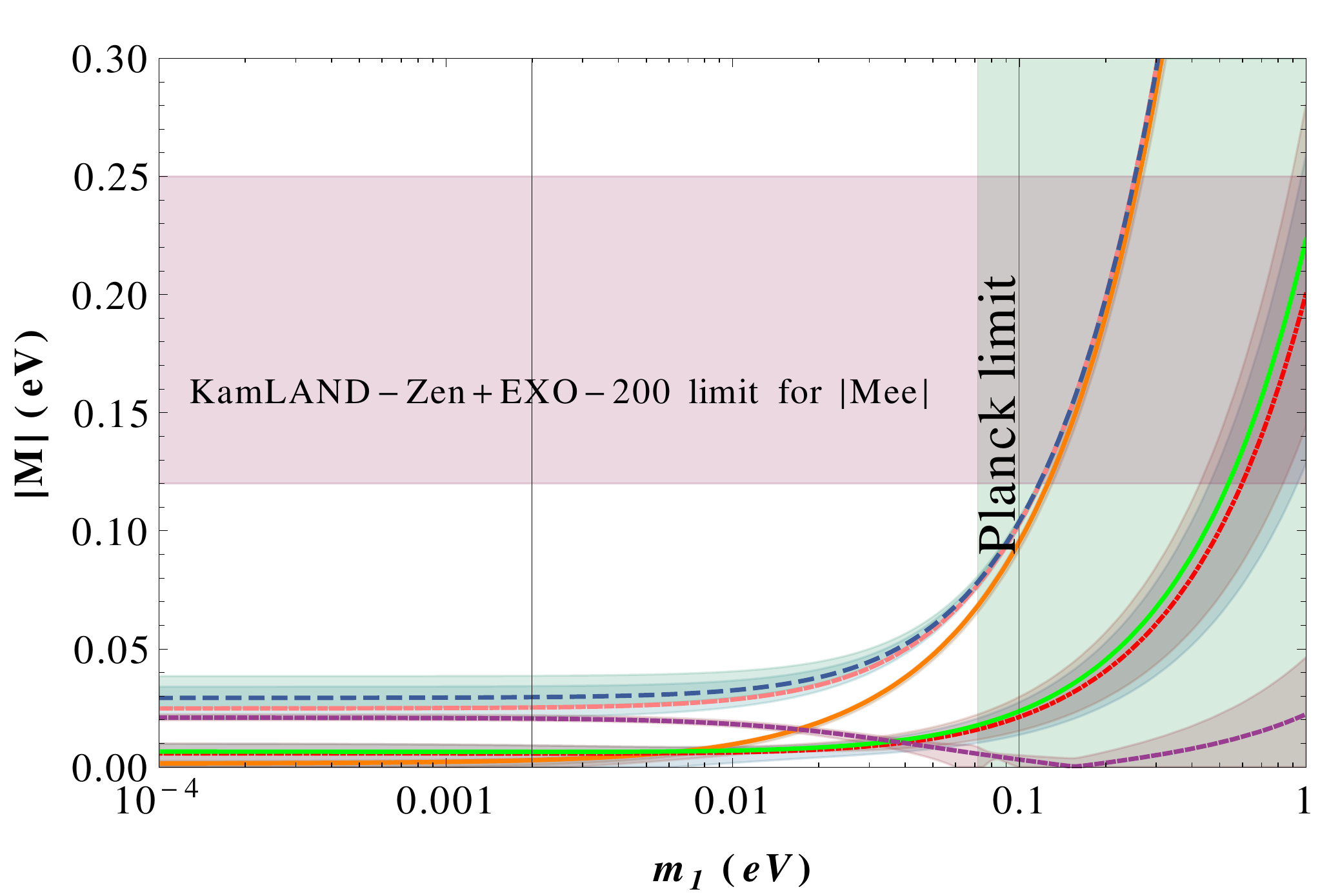}
   \includegraphics[width=.45\textwidth]{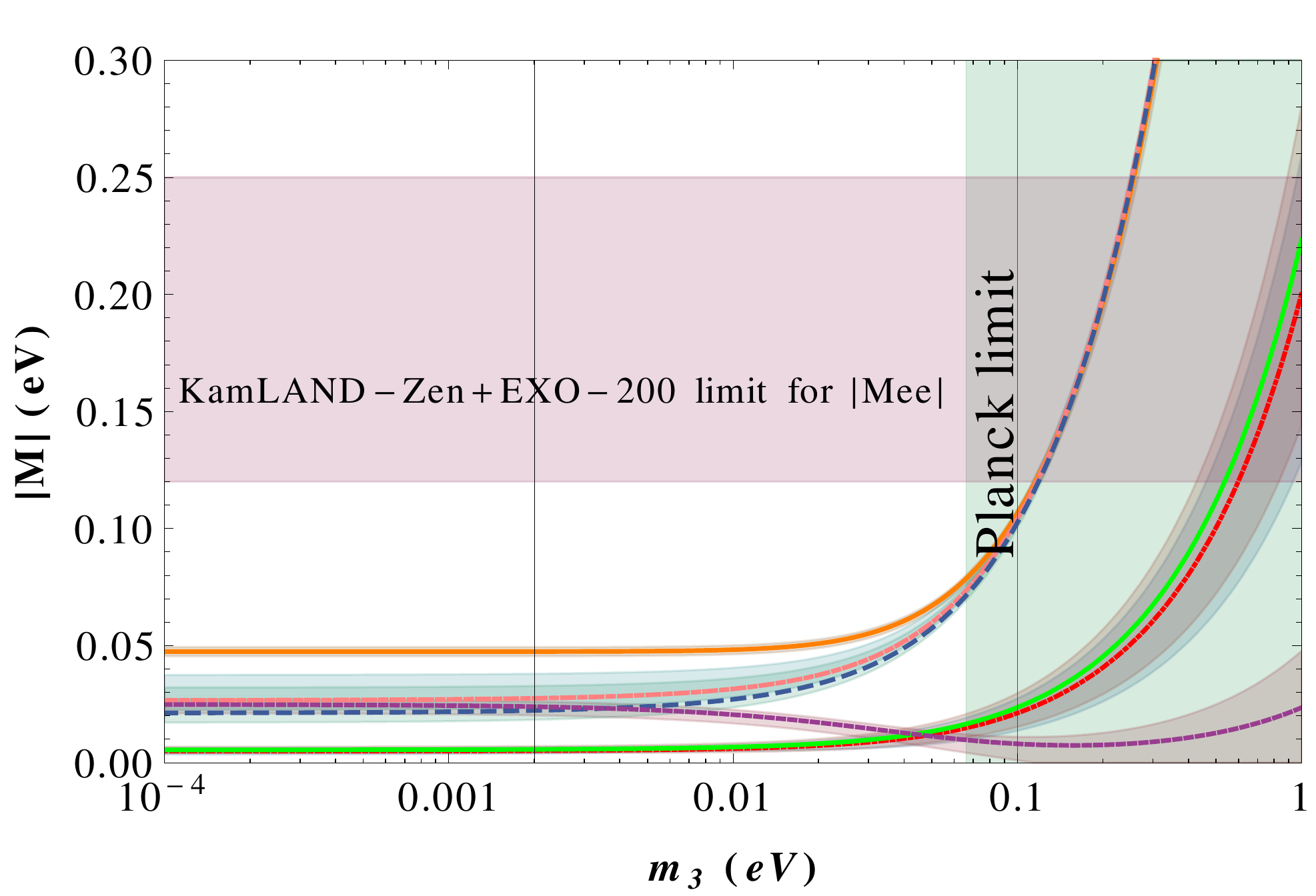}
           \caption{\label{fig:ii} The same as Figure~\ref{fig:i}, with extra demonstration of the exclusion areas setting by the non-oscillation data. The black vertical line distinguishes Region \uppercase\expandafter{\romannumeral1} from Region \uppercase\expandafter{\romannumeral2}. The red vertical line distinguishes Region \uppercase\expandafter{\romannumeral2} from Region \uppercase\expandafter{\romannumeral3}, and it is marked in comparison with the Planck limit.}
\end{minipage}
\end{figure*}

By far we only use the oscillation data, and it is helpful to get some constraints using non-oscillation data. We use the combined result from KamLAND-Zen and EXO-200 $\langle m_{\beta\beta}\rangle <(120-250)~\rm{meV}$~\cite{double beta}, and the Planck 2013 result for $\sum m_\nu<0.23~\rm{eV}$~\cite{planck}. We plot these constraints in Figure~\ref{fig:ii}.

We find that the cosmology limit has excluded Region \uppercase\expandafter{\romannumeral3}, while for the constraints from neutrinoless double beta decay experiments, a large part of Region \uppercase\expandafter{\romannumeral3} is excluded. Precision improvements in these experiments will help to narrow the allowed range for the lightest neutrino mass and distinguish the dominant structure eventually.

Given the value of $m_{\rm min}$, the dominant structure of the Majorana mass matrix can be read off the plots. Taking the normal ordering case as an illustration, we list the result of some $m_1$ values in Table~\ref{domstr} (we pick the crossing points and a random one in the regions divided by the crossing points). The dominant structure is parameterized by at most two parameters, with an intention to unveil its primary characteristics. For inverted ordering case, similar procedure can be taken.

Since the tribimaximal (TB) mixing is disfavored by the data, one may seek for other possible starting points for model building. We suggest a serious look at the Majorana mass matrix. Even if only the mixing is controlled by the symmetry, it does not mean a mixing pattern is actually valid for arbitrary masses. Because the mixing matrix is composed of eigenvectors of the mass matrix, it should have the one-to-one correspondence with the eigenvalues. On the other side, the values of mass and mixing parameters are unified in the mass matrix before it is diagonalized. Even if only the mixing is controlled by symmetry, starting from mass matrices would not hurt the conclusion since if one does it right the two kinds of models will coincide. The parameterizations in Table~\ref{domstr} can be served as starting points to build up a model in a bottom-up way.

\begin{table*}
\caption{Examples of the dominant structure of $\rm |M|$. }
\label{domstr}
\begin{tabular*}{\textwidth}{@{\extracolsep{\fill}}lrrrrl@{}}
\hline
$m_1 (\rm{eV})$ & Numerical results & Parametrization & Values of parameters  \\
\hline
0.001 &
\multirow{3}{*}{
  $\left(
   \begin{array}{ccc}
   0.0022 & 0.0057 & 0.0064 \\
   .      & 0.0249 & 0.0206 \\
   .      & .      &0.0294  \\
      \end{array}
   \right)$} &
\multirow{3}{*}{
  $\left(
   \begin{array}{ccc}
   a/3 & a & a \\
   .      & 4a & 3a \\
   .      & .      &5a  \\
      \end{array}
   \right)$}  &
   $a\sim0.006$\\
&&&\\
&&&\\
&&&\\
0.006 &
\multirow{3}{*}{
  $\left(
   \begin{array}{ccc}
   0.0058 & 0.0058 & 0.0066 \\
   .      & 0.0265 & 0.0191 \\
   .      & .      &0.0307  \\
      \end{array}
   \right)$} &
\multirow{3}{*}{
  $\left(
   \begin{array}{ccc}
   a   & a  & a  \\
   .   & 4a & 3a \\
   .   & .  & 5a \\
      \end{array}
   \right)$}  &
   $a\sim0.006$\\
$|\rm M_{ee}|=|\rm M_{e\mu}|\simeq|\rm M_{e\tau}|$&&&\\
&&&\\
&&&\\
0.01 &
\multirow{3}{*}{
  $\left(
   \begin{array}{ccc}
   0.0096 & 0.0062 & 0.0069 \\
   .      & 0.0285 & 0.0174 \\
   .      & .      & 0.0325 \\
      \end{array}
   \right)$} &
\multirow{3}{*}{
  $\left(
   \begin{array}{ccc}
   b   & a  & a  \\
   .   & 3a+b & 3a \\
   .   & .  & 2a+b \\
      \end{array}
   \right)$}  &
   $a\sim0.006$\\
&&&$b\sim0.01$\\
&&&\\
&&&\\
0.016 &
\multirow{3}{*}{
  $\left(
   \begin{array}{ccc}
   0.0150 & 0.0068 & 0.0077 \\
   .      & 0.0320 & 0.0150 \\
   .      & .      & 0.0355 \\
      \end{array}
   \right)$} &
\multirow{3}{*}{
  $\left(
   \begin{array}{ccc}
   a   & a/2  & a/2  \\
   .   & 2a & a \\
   .   & .  & 2a \\
      \end{array}
   \right)$}  &
   $a\sim0.016$\\
$|\rm M_{ee}|=|\rm M_{\mu\tau}|$ &&&\\
&&&\\
&&&\\
0.02 &
\multirow{3}{*}{
  $\left(
   \begin{array}{ccc}
   0.0189 & 0.0074 & 0.0083 \\
   .      & 0.0347 & 0.0134 \\
   .      & .      & 0.0379 \\
      \end{array}
   \right)$} &
\multirow{3}{*}{
  $\left(
   \begin{array}{ccc}
   3a-b & a  & a\\
   .   & 5a-b & a+b \\
   .   & .  & 4a+b \\
      \end{array}
   \right)$}  &
   $a\sim0.008$\\
&&&$b\sim0.005$\\
&&&\\
&&&\\
0.03 &
\multirow{3}{*}{
  $\left(
   \begin{array}{ccc}
   0.0284 & 0.0088 & 0.0098 \\
   .      & 0.0419 & 0.0100 \\
   .      & .      & 0.0445 \\
      \end{array}
   \right)$} &
\multirow{3}{*}{
  $\left(
   \begin{array}{ccc}
   3a  & a  & a\\
   .   & 4a & a\\
   .   & .  & 4a\\
      \end{array}
   \right)$}  &
   $a\sim0.01$\\
$|\rm M_{\mu\tau}|\simeq|\rm M_{e\tau}|\simeq|\rm M_{e\mu}|$&&&\\
&&&\\
&&&\\
0.06 &
\multirow{3}{*}{
  $\left(
   \begin{array}{ccc}
   0.0570 & 0.0138 & 0.0154 \\
   .      & 0.0667 & 0.0025 \\
   .      & .      & 0.0682 \\
      \end{array}
   \right)$} &
\multirow{3}{*}{
  $\left(
   \begin{array}{ccc}
   a-4b & 5b  & 5b\\
   .   & a-b & b \\
   .   & .  & a-b \\
      \end{array}
   \right)$}  &
   $a\sim0.07$\\
&&&$b\sim0.003$\\
&&&\\
&&&\\
\hline
\end{tabular*}
\end{table*}
\section{Discussion}
\label{sec:discuss}
\subsection{Regional characteristics}

In Region \uppercase\expandafter{\romannumeral1} of the normal ordering case, the relative magnitudes determined by the best fit values of the inputting parameters are found to be
\begin{eqnarray}
|\rm M_{\tau\tau}|>|\rm M_{\mu\mu}|>|\rm M_{\mu\tau}|>|\rm M_{e\tau}|>|\rm M_{e\mu}|>|\rm M_{ee}|.\label{eqn:no}
\end{eqnarray}

The first three are of $\mathcal{O}(10^{-2})\rm~eV$, while the latter three are of $\mathcal{O}(10^{-3})\rm~eV$. The two groups are distinguishable in $3 \sigma$ range. We find $|\rm M_{\mu\mu}|\simeq|\rm M_{\tau\tau}|$ and the curves determined by the best fit values of the inputting parameters lie out of the $3 \sigma$ range of $|\rm M_{\mu\tau}|$. Similarly, $|\rm M_{e\tau}|\simeq|\rm M_{e\mu}|$ and the curves determined by the best fit inputting parameters lie out of the $3 \sigma$ range of $|\rm M_{ee}|$. Thus, the $\mu$-$\tau$ exchange symmetry is recognized.

It is well known that the mass matrix of the form
\begin{eqnarray}
\rm M_{BM}=\left(
   \begin{array}{ccc}
   x & y  & y\\
   .   & z & x-z \\
   .   & .  & z \\
      \end{array}
   \right),\quad
\rm M_{TB}=\left(
   \begin{array}{ccc}
   x & y  & y\\
   .   & x+v & y-v \\
   .   & .  & x+v \\
      \end{array}
   \right)  \label{eqn:mtbbm}
\end{eqnarray}
can be diagonalized by the bimaximal (BM)~\cite{BM1,BM2,BM3} and the tribimaximal (TB)~\cite{TB1,TB2} mixing matrix correspondingly. From Eq.~(\ref{eqn:mtbbm}), the following relations among the mass matrix entries can be found:
\begin{eqnarray}
\rm M_{ee}=\rm M_{\mu\mu}+\rm M_{\mu\tau},\quad from~\rm{BM};\label{eqn:BM}\\
\rm M_{ee}+\rm M_{e\mu}=\rm M_{\mu\mu}+\rm M_{\mu\tau},\quad from~\rm{TB}.\label{eqn:TB}
\end{eqnarray}

As seen from the magnitudes grouping, both of the above relations cannot be satisfied in Region \uppercase\expandafter{\romannumeral1} of the normal ordering. Notice that the TB mixing and the BM mixing also indicate $\rm M_{e\mu}=\rm M_{e\tau},~\rm M_{\mu\mu}=\rm M_{\tau\tau}$, which are the relations produced by the $\mu$-$\tau$ exchange symmetry and we discuss them separately.

For the inverted ordering in Region \uppercase\expandafter{\romannumeral1}, we have
\begin{eqnarray}
|\rm M_{ee}|>|\rm M_{\mu\mu}|>|\rm M_{\mu\tau}|>|\rm M_{\tau\tau}|>|\rm M_{e\tau}|>|\rm M_{e\mu}|.\label{eqn:io}
\end{eqnarray}
The first four are of $\mathcal{O}(10^{-2})\rm~eV$, while the latter two are of $\mathcal{O}(10^{-3})\rm~eV$. The relations in eqns.~\ref{eqn:BM}~and~\ref{eqn:TB} are satisfied approximately. We emphasize here that since eqns.~\ref{eqn:BM}~and~\ref{eqn:TB} alone cannot determine the mixing pattern ($\rm M_{e\mu}=\rm M_{e\tau},~\rm M_{\mu\mu}=\rm M_{\tau\tau}$ are needed), and these relations are not satisfied in the same region at the same degree, we cannot say that the TB or the BM mixing is realized in this region.

\begin{figure*}
\begin{minipage}{\textwidth}
   \centering
  \includegraphics[width=.45\textwidth]{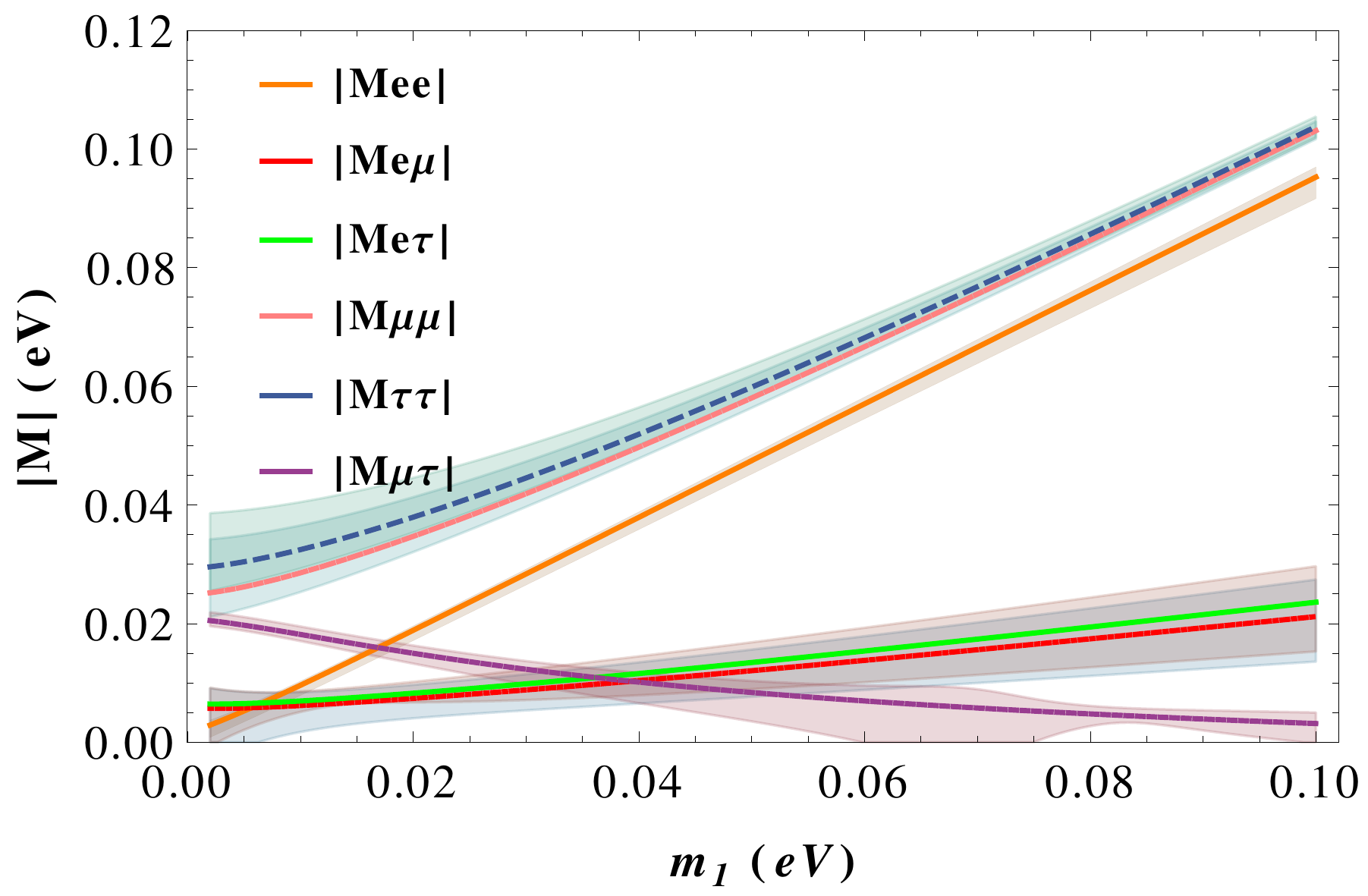}
  \includegraphics[width=.45\textwidth]{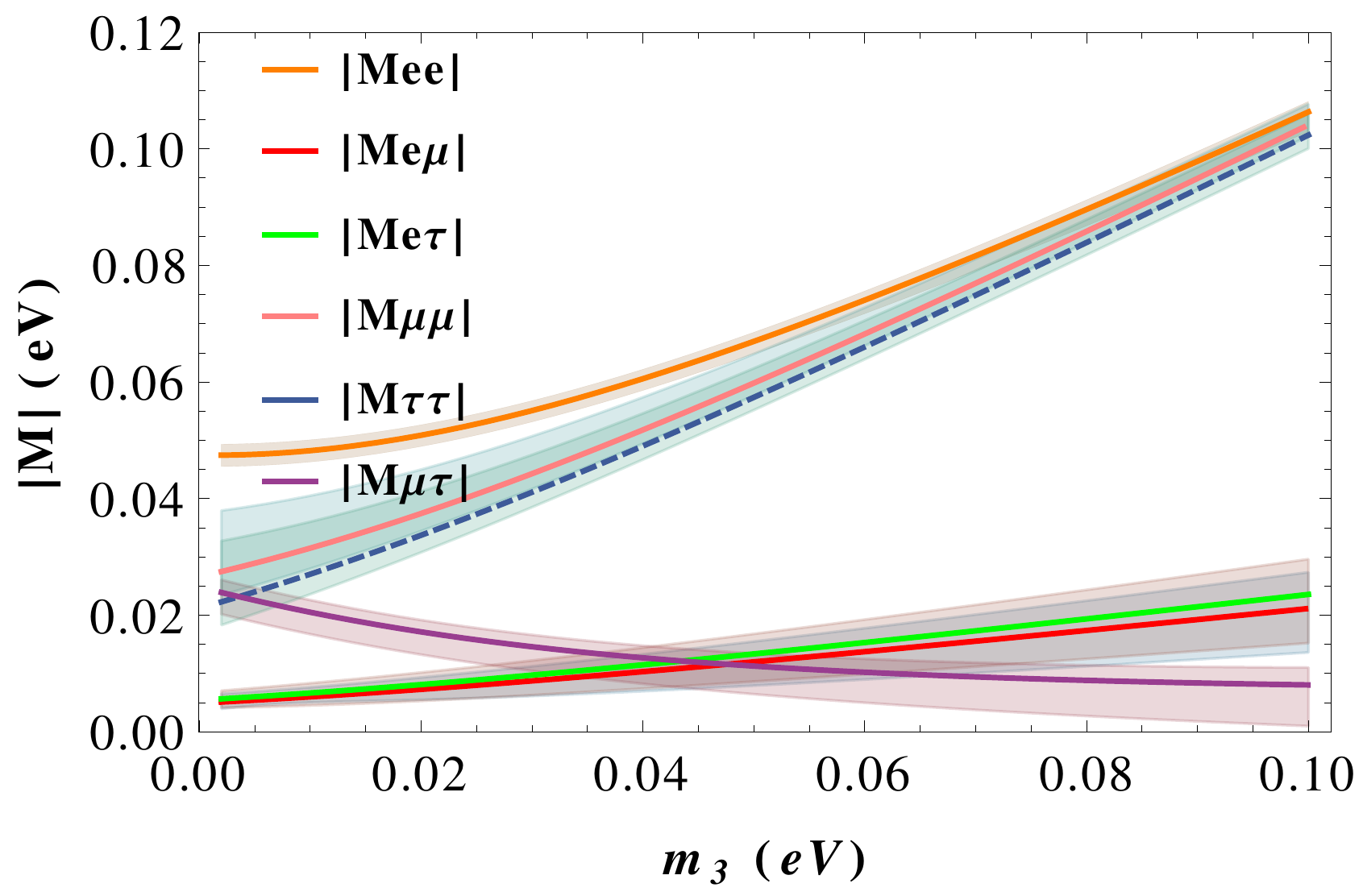}
           \caption{$|\rm M_{\alpha\beta}|$ as a function of the lightest neutrino mass in Region \uppercase\expandafter{\romannumeral2}. The left one corresponds to the normal ordering, and the right one corresponds to the inverted ordering.}
  \label{fig:region2}
\end{minipage}
\end{figure*}

Figure~\ref{fig:region2} is offered to see Region \uppercase\expandafter{\romannumeral2} more clearly.

We see that in the normal ordering, the BM relation~\ref{eqn:BM} still cannot be satisfied, while the TB relation~\ref{eqn:TB} can be satisfied approximately in $0.05~\rm{eV}<\emph{m}_1<0.08~\rm{eV}$. In the inverted ordering, both of the relations can be satisfied approximately.

Since Region \uppercase\expandafter{\romannumeral3} is excluded by the cosmology limit, we only make a short comment. We also observe the $\mu$-$\tau$ exchange symmetry in both orderings. Besides, there are three groups with distinguishable magnitudes, that is,

\begin{enumerate}
\item $|\rm M_{ee}|,|\rm M_{\mu\mu}|,|\rm M_{\tau\tau}|$, which have the largest magnitudes in given $m_{\rm{min}}$.\\
\item $|\rm M_{e\mu}|,|\rm M_{e\tau}|$ with an intermediate magnitude.\\
\item $|\rm M_{\mu\tau}|$ with the smallest magnitude.
\end{enumerate}

To sum up, we observe an approximating $\mu$-$\tau$ exchange symmetry in all the three regions of both the normal ordering and the inverted ordering. We also observe a ``grouping effect'' in Region \uppercase\expandafter{\romannumeral1} and Region \uppercase\expandafter{\romannumeral3}. Different groups are distinguishable at a $3\sigma$ level.

\subsection{$\mu$-$\tau$ exchange symmetry}
As is mentioned, we observe $|\rm M_{e\mu}|\simeq|\rm M_{e\tau}|,\quad |\rm M_{\mu\mu}|\simeq|\rm M_{\tau\tau}|$ in all the regions of both orderings. These relations can result from the $\mu$-$\tau$ exchange symmetry.

By definition, the $\mu$-$\tau$ exchange symmetry is the invariance of the Lagrangian under an exchange of $\nu_{\mu}$ with $\nu_{\tau}$. In the Majorana mass term, it means
\begin{eqnarray}
\mathcal{L}&=&\frac{1}{2}(\bar\nu_{eL},\bar\nu_{\mu L},\bar\nu_{\tau L}) \rm{M} (\nu_{eL}^c,\nu_{\mu L}^c,\nu_{\tau L}^c)^T+H.c.,\nonumber\\
&=&\frac{1}{2}(\bar\nu_{eL},\bar\nu_{\tau L},\bar\nu_{\mu L}) \rm{M} (\nu_{eL}^c,\nu_{\tau L}^c,\nu_{\mu L}^c)^T+H.c..
\end{eqnarray}
Immediately, one arrives at
\begin{eqnarray}
\rm M_{e\mu}=\rm M_{e\tau},\quad \rm M_{\mu\mu}=\rm M_{\tau\tau}.\label{eqn:mutao}
\end{eqnarray}
An exchange of $\nu_{\mu}$ with $\nu_{\tau}$ in the mass term can be performed with the following transformation matrix:
\begin{eqnarray}
\rm A_{23}=\left(
  \begin{array}{ccc}
    1  & 0  & 0 \\
    0  & 0  & 1 \\
    0  & 1  & 0 \\
  \end{array}\right).
\end{eqnarray}
The invariance of the mass term means $\rm A_{23} \rm M \rm A_{23}=\rm M $. Since the matrix $\rm A$ generates a $Z_2$ symmetry, one may also consider the invariance corresponding to the eigenvalue of $-1$, that is, $\rm A_{23} \rm M \rm A_{23}=-\rm M $.

In the flavor basis, the Majorana matrix is diagonalized by the PMNS matrix, that is, $\rm U^T \rm M \rm U= \rm Diag\{m_1,m_2,m_3\}$. Thus, we get the $\mu$-$\tau$ exchange symmetry relation in terms of the mixing matrix entries, that is,
\begin{eqnarray}
\rm U_{\mu i}=\rm U_{\tau i},
\end{eqnarray}
which is necessary and sufficient. Regarding constraints on the mixing angles and the phase, when $|\rm U_{\mu i}|=|\rm U_{\tau i}|$, one gets one of the following two consequences:
\begin{enumerate}
\item $\theta_{23}=\pi/4,\quad \delta=\pm \pi/2$;
\item $\theta_{23}=\pi/4,\quad \theta_{13}=0$.
\end{enumerate}
Notice that both of the above two results are necessary but not sufficient conditions for the $\mu$-$\tau$ exchange symmetry. Notice also the current data indicate a deviation from $\theta_{23}=\pi/4$, which also means the $\mu$-$\tau$ exchange symmetry is an approximate symmetry, as can be seen from the figures in this paper directly.

This symmetry has been extensively studied~\cite{23symmetry1,23symmetry2,23symmetry3,23symmetry4,23symmetry5,23symmetry6,23symmetry7,23symmetry8,23symmetry9,23symmetry10,23symmetry11,23symmetry12,23symmetry13}. After the measurements of $\theta_{13}$, dedicated investigations of the $\mu$-$\tau$ exchange symmetry can be found in ref.~\cite{Grimus:2012hu,Gupta:2013it}. It is also shown recently that the $\mu$-$\tau$ exchange symmetry, in combination with other inputs, can lead to a new mixing pattern which is simple and elegant~\cite{Qu2013}.

\subsection{Texture non-zeros}
As an application to our results, we display the texture non-zeros in Table~\ref{nonzero}. By texture non-zeros, we refer to the mass matrix elements $|\rm M|_{\alpha\beta}$ that are non-zero at $3 \sigma$ level indicated by current experimental data. The only reason we use non-zeros rather than zeros is that categorizing in this way is much simpler. It is simple in the sense that each region has only one non-zero texture. For the one non-zero texture, specifying the value of $m_{\rm min}$ in more detail will unveil the texture zeros. It is equivalent to finding the possible texture zeros. For example, we find that the textures in Region \uppercase\expandafter{\romannumeral1} of normal ordering is compatible with the two-zero textures $\rm A_1$ and $\rm A_2$ in ref.~\cite{Frampton:2002yf}.

Since it is the invariance of the Majorana mass matrix under some transformations that exhibits the residual symmetry, it is important to know the dominant structure of the Majorana mass matrix. Our investigation provides an intuitive display of the structure. Although, in general, a dominant structure is not distinguishable due to the errors of the parameters, one can still find some information on it. For example, for the group with a large magnitude, the non-zero is established at $3 \sigma$ level, which allows for comparison with texture-zero models.

\begin{table*}
\caption{Texture non-zeros of the Majorana mass matrix $(\alpha_{21}, \alpha_{31})=(0,0)$.}
\label{nonzero}
\begin{tabular*}{\textwidth}{@{\extracolsep{\fill}}lrrrrl@{}}
\hline
 & Normal ordering & Inverted ordering  \\
\hline
Region \uppercase\expandafter{\romannumeral1} &
  \multirow{3}{*}{
  $\left(
      \begin{array}{ccc}
   \square & \square & \square \\
     .     & \star   & \star \\
     .     & .       & \star  \\
      \end{array}
   \right)$}\footnote{``$\star$" stands for the $3 \sigma$ non-zero matrix element, and ``$\square$" stands for possible zero matrix element at $3 \sigma$ level. }
   &
   \multirow{3}{*}{
     $\left(
         \begin{array}{ccc}
      \star & \square & \square \\
        .     & \star   & \star \\
        .     & .       & \star  \\
         \end{array}
      \right)$} \\
   &&\\
   &&\\
\hline
Region \uppercase\expandafter{\romannumeral2} &
  \multirow{3}{*}{
  $\left(
      \begin{array}{ccc}
   \square & \square & \square \\
     .     & \star   & \square \\
     .     & .       & \star  \\
      \end{array}
   \right)$}
   &
   \multirow{3}{*}{
     $\left(
         \begin{array}{ccc}
        \star & \square & \square \\
        .     & \star   & \square \\
        .     & .       & \star  \\
         \end{array}
      \right)$} \\
   &&\\
   &&\\
\hline
Region \uppercase\expandafter{\romannumeral3} &
  \multirow{3}{*}{
  $\left(
      \begin{array}{ccc}
     \star & \star   & \star\\
     .     & \star   & \square \\
     .     & .       & \star  \\
      \end{array}
   \right)$}
   &
   \multirow{3}{*}{
     $\left(
         \begin{array}{ccc}
      \star   & \star   & \star \\
        .     & \star   & \square \\
        .     & .       & \star  \\
         \end{array}
      \right)$} \\
   &&\\
   &&\\
\hline
\end{tabular*}
\end{table*}

\subsection{effects of the Majorana phases}
\begin{figure*}
\begin{minipage}{\textwidth}
   \centering
  \includegraphics[width=.45\textwidth]{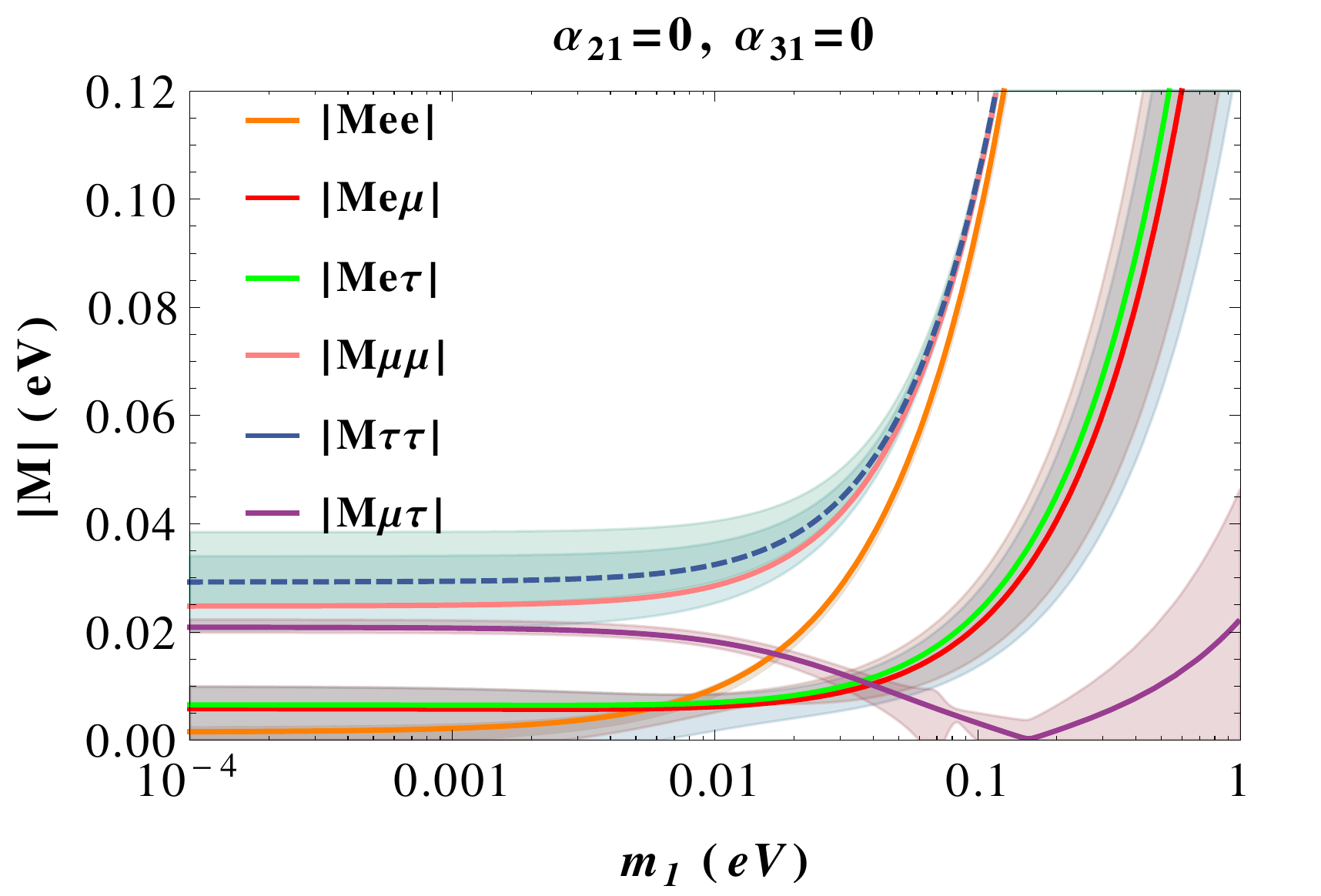}
  \includegraphics[width=.45\textwidth]{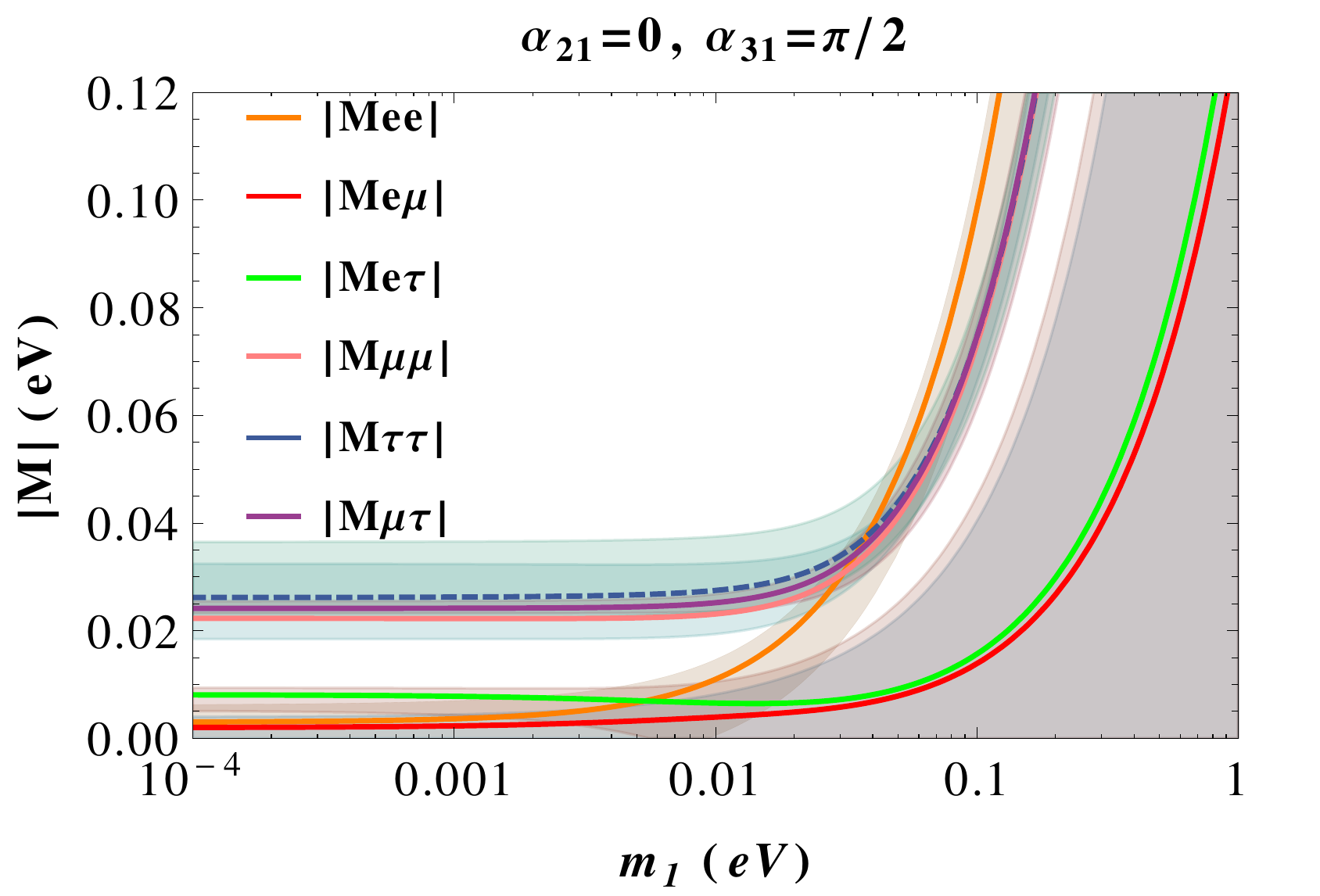}\\
  \includegraphics[width=.45\textwidth]{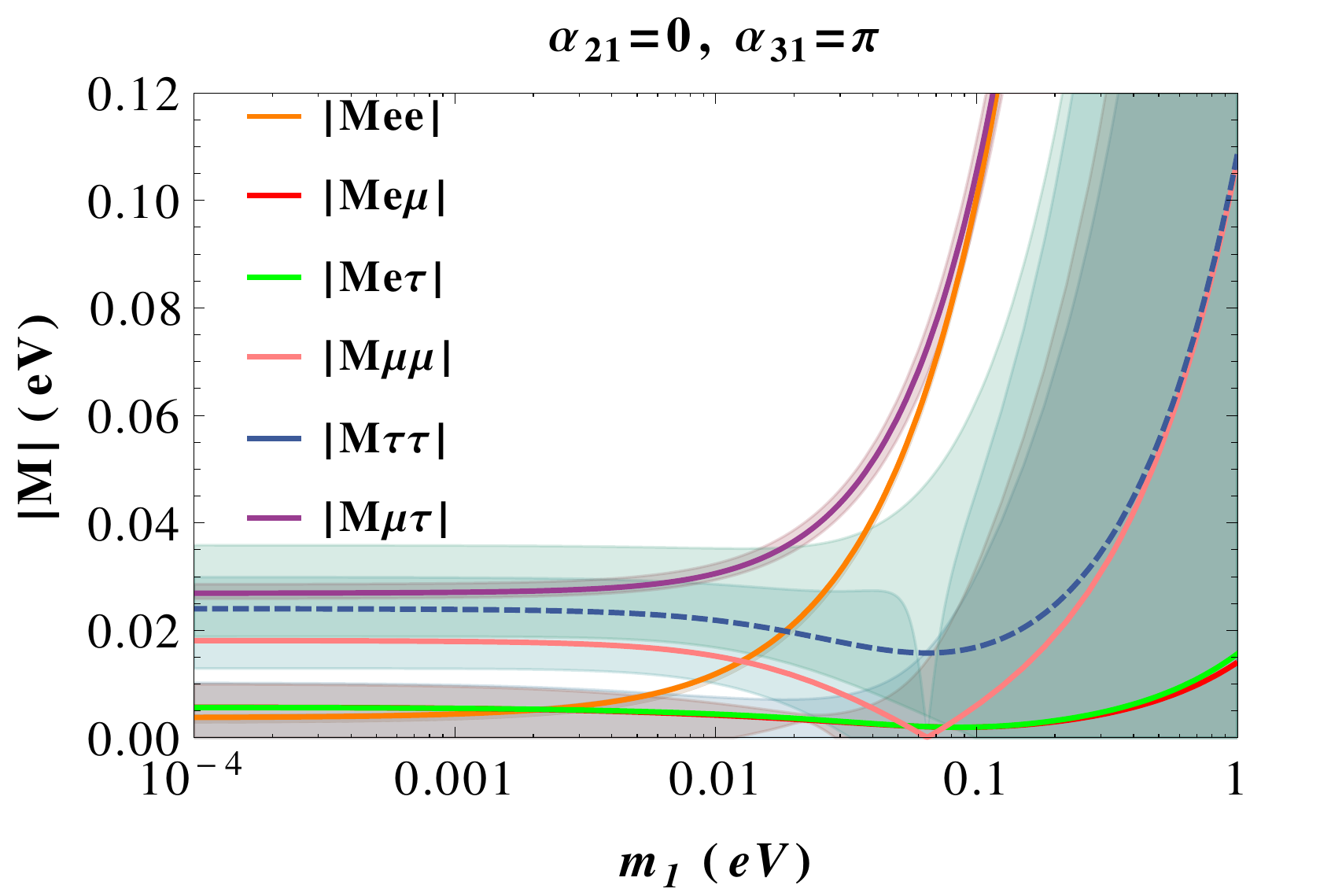}
  \includegraphics[width=.45\textwidth]{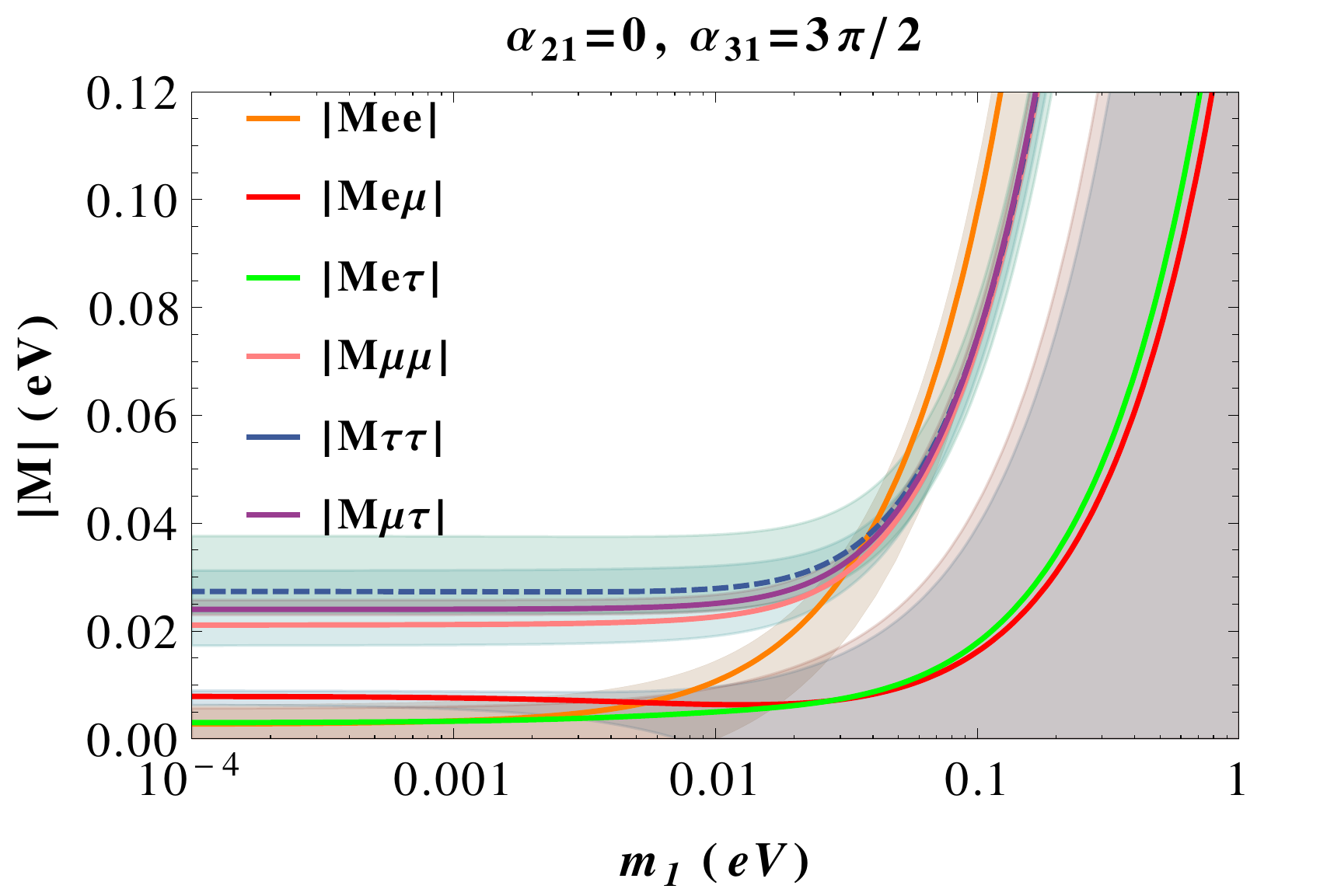}\\
  \includegraphics[width=.45\textwidth]{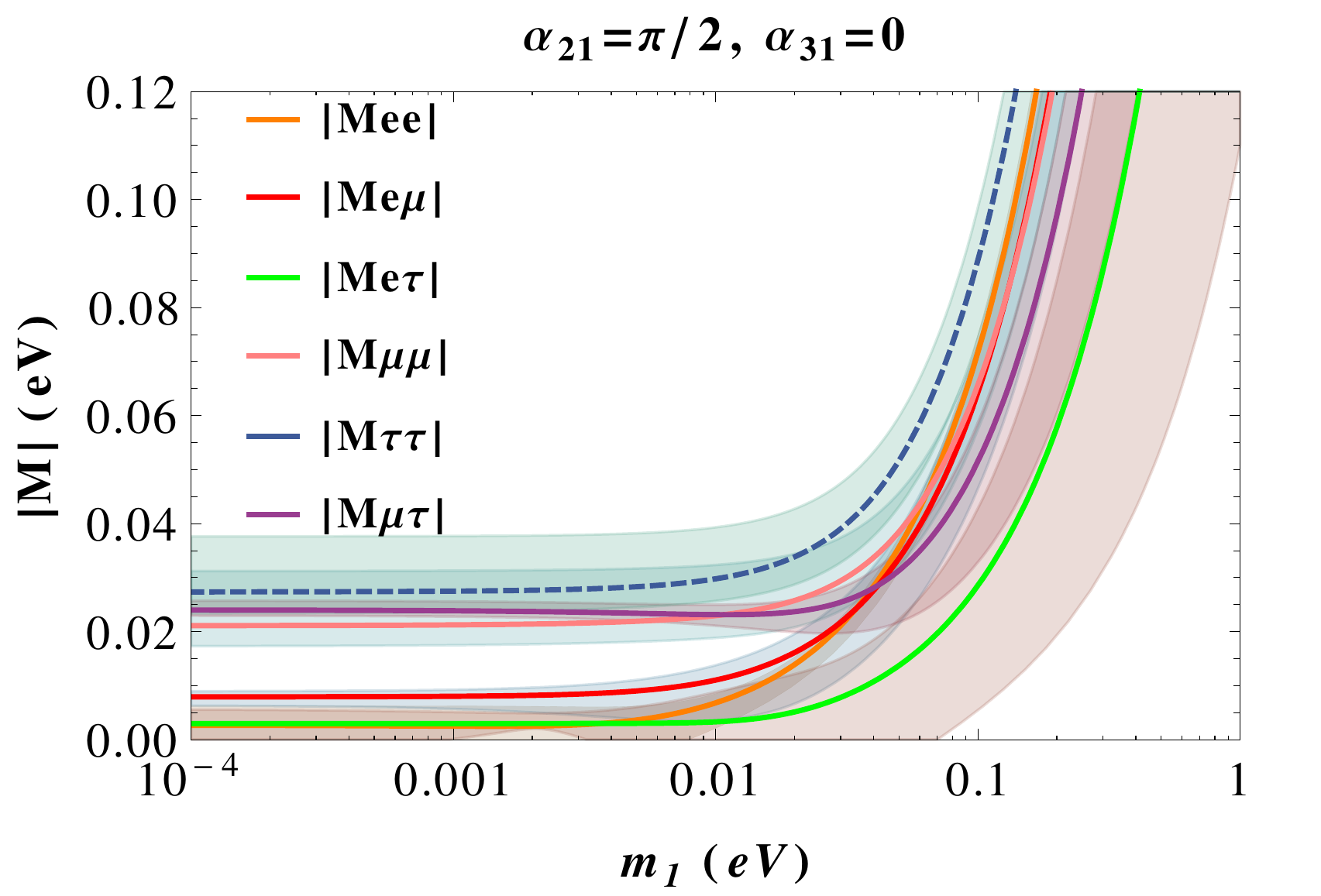}
  \includegraphics[width=.45\textwidth]{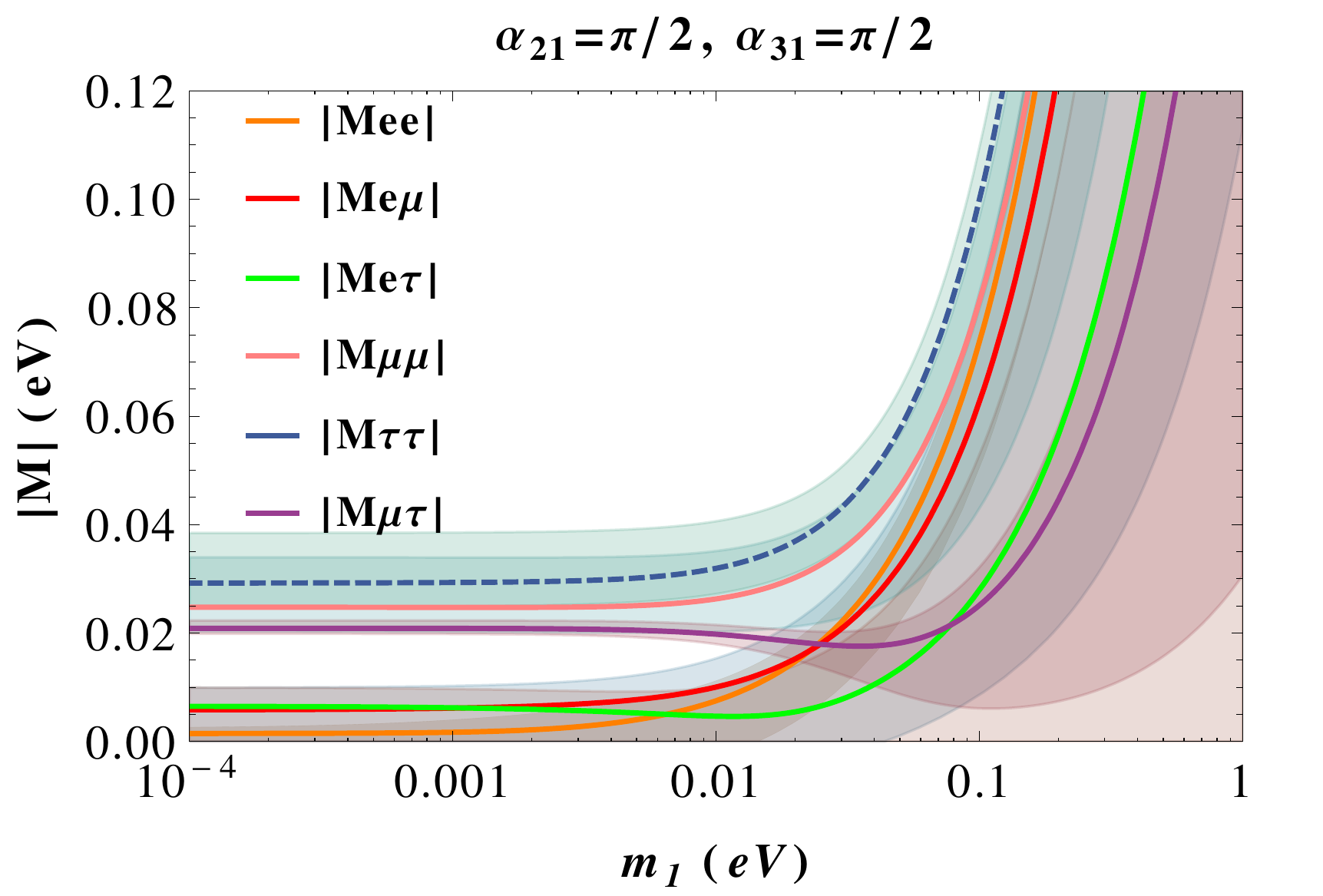}\\
  \includegraphics[width=.45\textwidth]{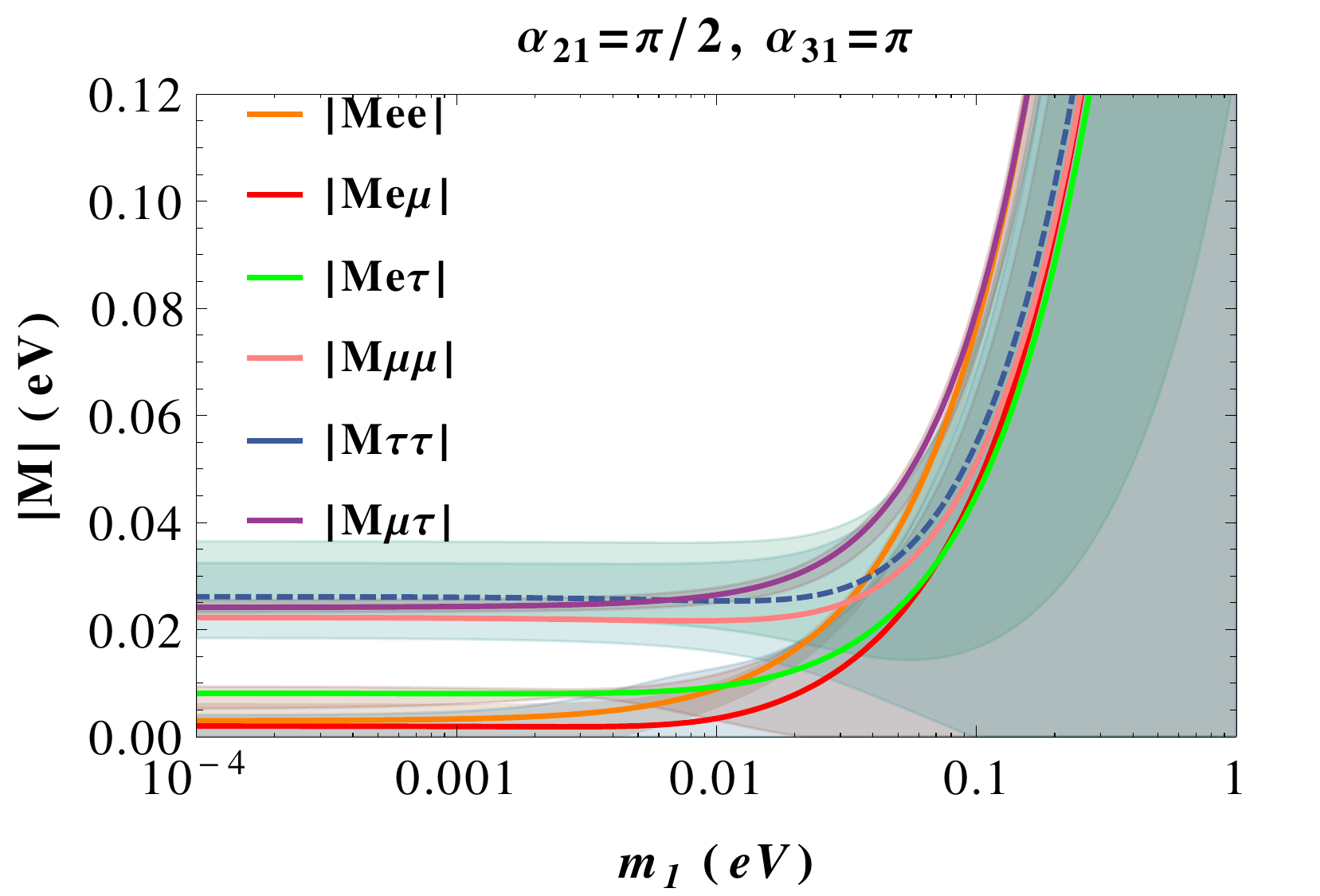}
  \includegraphics[width=.45\textwidth]{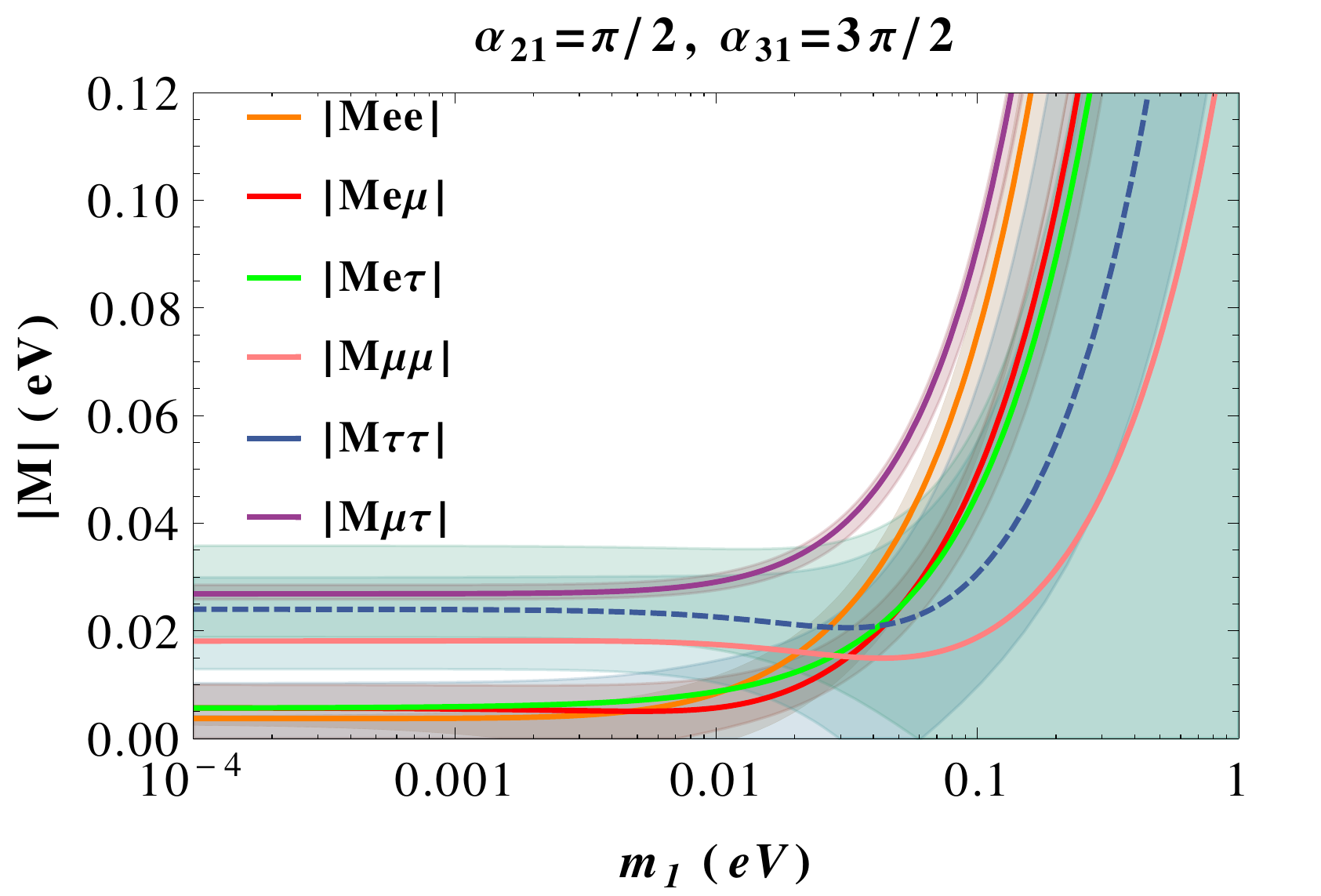}
         \caption{$|\rm M_{\alpha\beta}|$ as a function of the lightest neutrino mass with $(\alpha_{21}, \alpha_{31})$ values equal to $(0,0)$, $(0,\frac{\pi}{2})$, $(0,\pi)$, $(0,\frac{3\pi}{2})$, $(\frac{\pi}{2},0)$ , $(\frac{\pi}{2},\frac{\pi}{2})$, $(\frac{\pi}{2},\pi)$, $(\frac{\pi}{2},\frac{3\pi}{2})$ in the case of the normal ordering.}
  \label{fig:alpha-no1}
\end{minipage}
\end{figure*}

To see the effects of the Majorana phases on the structure of neutrino mass, we illustrate the normal ordering results with the Majorana phases $(\alpha_{21},\alpha_{31})$ setting to pairs of special values in Figures~\ref{fig:alpha-no1}~and~\ref{fig:alpha-no2}. The case of the inverted ordering can be found in Section~\ref{sec:io} of the Appendix. Regarding the texture non-zeros, we list the case for $(\alpha_{21}, \alpha_{31})$ values equal to $(0,\frac{\pi}{2})$ in Table~\ref{nonzero2} for illustration. We observe the changes of the texture non-zeros as anticipated.

\begin{table*}
\caption{Texture non-zeros of the Majorana mass matrix $(\alpha_{21}, \alpha_{31})=(0,\frac{\pi}{2})$.}
\label{nonzero2}
\begin{tabular*}{\textwidth}{@{\extracolsep{\fill}}lrrrrl@{}}
\hline
 & Normal ordering & Inverted ordering  \\
\hline
Region \uppercase\expandafter{\romannumeral1} &
  \multirow{3}{*}{
  $\left(
      \begin{array}{ccc}
   \square & \square & \square \\
     .     & \star   & \star \\
     .     & .       & \star  \\
      \end{array}
   \right)$}\footnote{``$\star$" stands for the $3 \sigma$ non-zero matrix element, and ``$\square$" stands for possible zero matrix element at $3 \sigma$ level. }
   &
   \multirow{3}{*}{
     $\left(
         \begin{array}{ccc}
      \star & \square & \square \\
        .     & \star   & \star \\
        .     & .       & \star  \\
         \end{array}
      \right)$} \\
   &&\\
   &&\\
\hline
Region \uppercase\expandafter{\romannumeral2} &
  \multirow{3}{*}{
  $\left(
      \begin{array}{ccc}
   \square & \square & \square \\
     .     & \star   & \star \\
     .     & .       & \star  \\
      \end{array}
   \right)$}
   &
   \multirow{3}{*}{
     $\left(
         \begin{array}{ccc}
        \star & \square & \square \\
        .     & \star   & \star \\
        .     & .       & \star  \\
         \end{array}
      \right)$} \\
   &&\\
   &&\\
\hline
Region \uppercase\expandafter{\romannumeral3} &
  \multirow{3}{*}{
  $\left(
      \begin{array}{ccc}
     \star & \square & \square\\
     .     & \star   & \star \\
     .     & .       & \star  \\
      \end{array}
   \right)$}
   &
   \multirow{3}{*}{
     $\left(
         \begin{array}{ccc}
        \star & \square & \square\\
        .     & \star   & \star \\
        .     & .       & \star  \\
         \end{array}
      \right)$} \\
   &&\\
   &&\\
\hline
\end{tabular*}
\end{table*}

\begin{figure*}
\begin{minipage}{\textwidth}
   \centering
  \includegraphics[width=.45\textwidth]{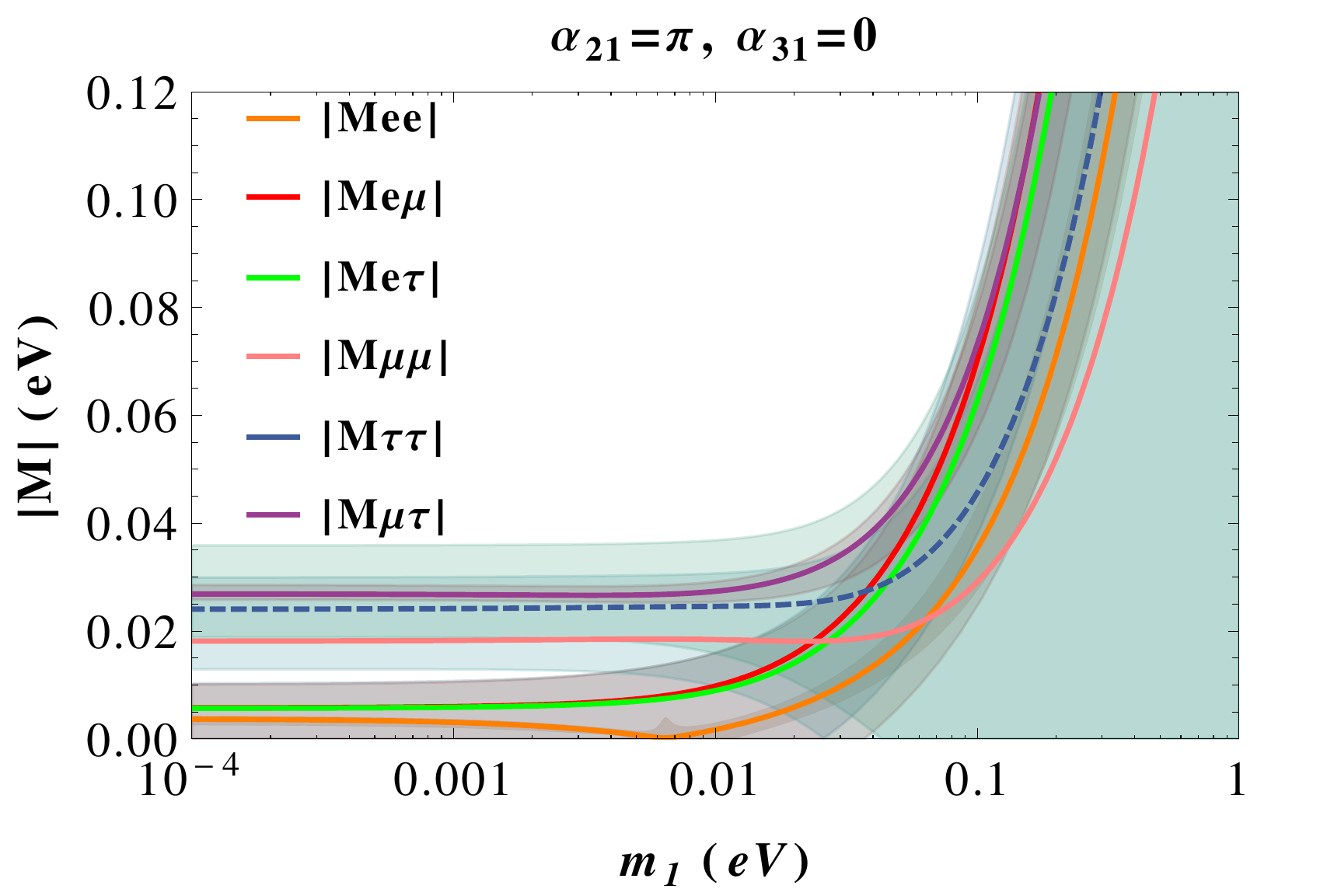}
   \includegraphics[width=.45\textwidth]{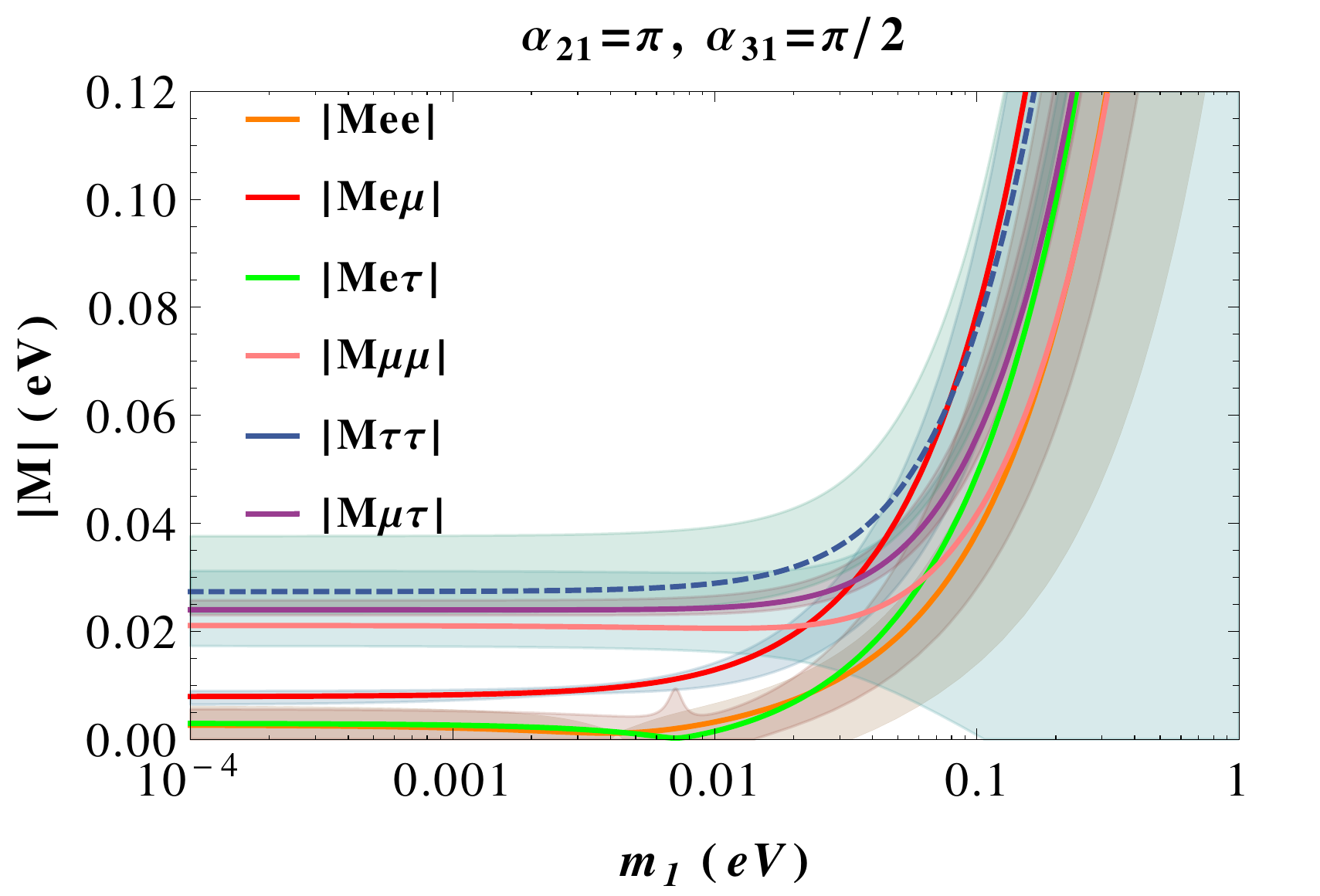}\\
   \includegraphics[width=.45\textwidth]{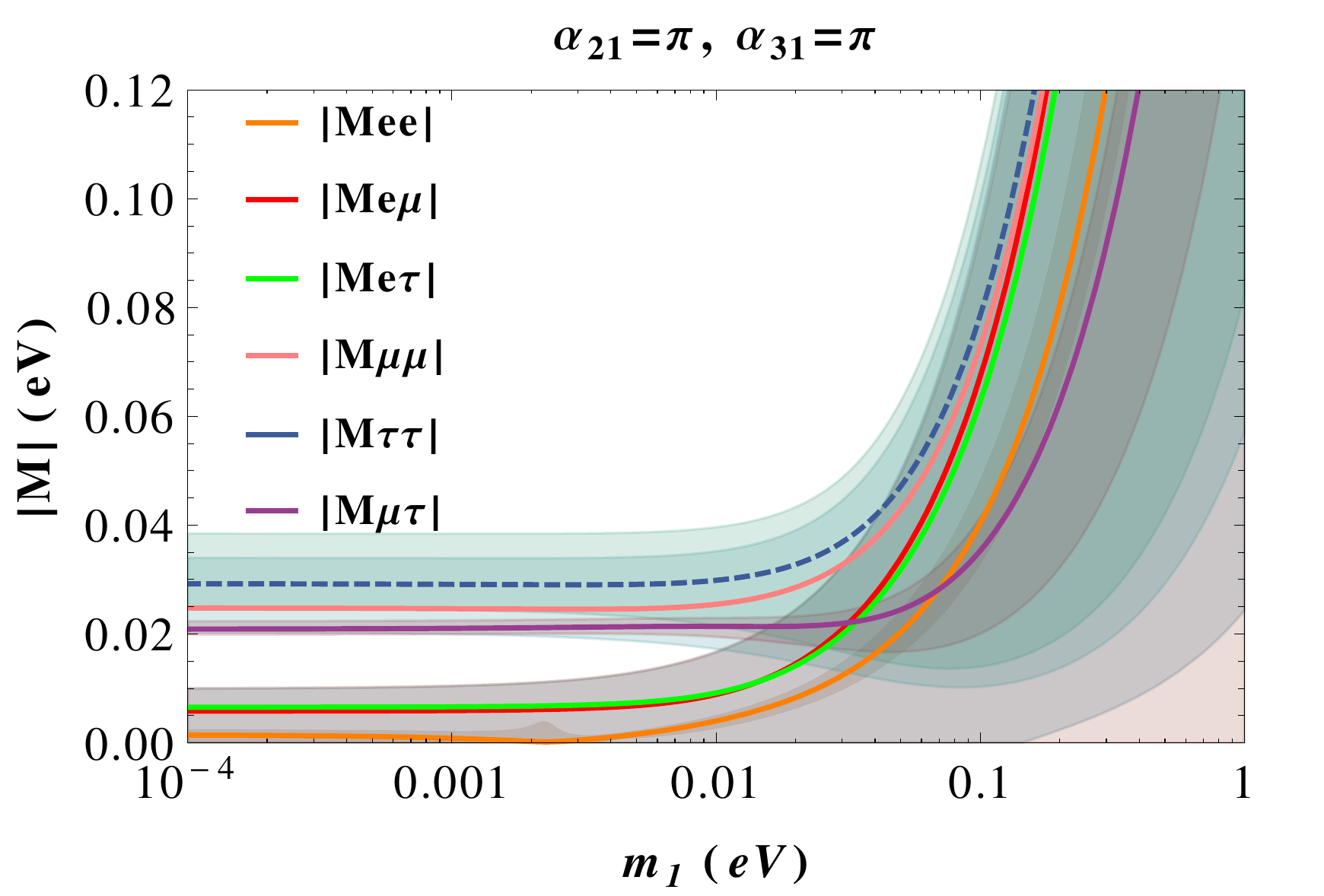}
   \includegraphics[width=.45\textwidth]{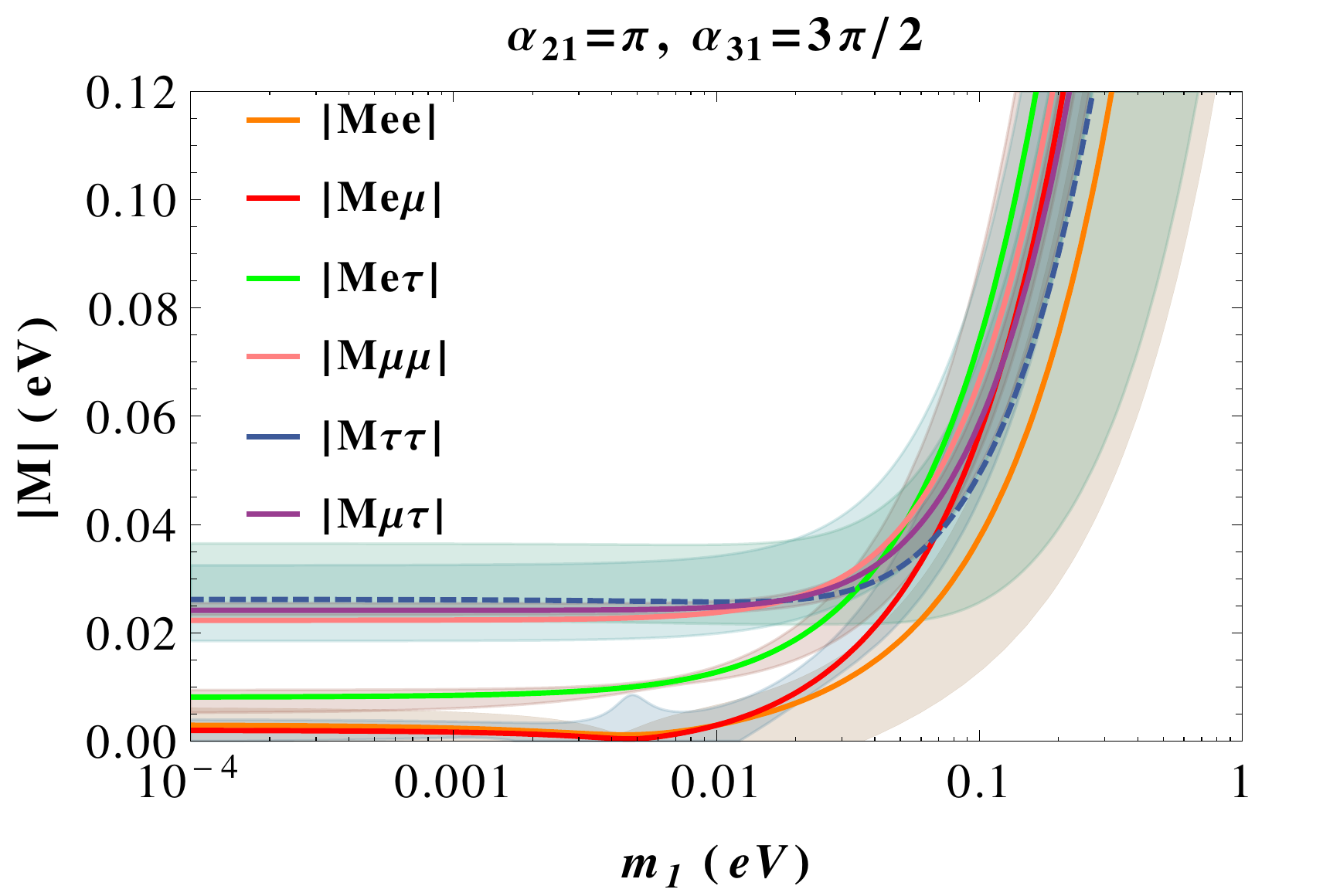}\\
   \includegraphics[width=.45\textwidth]{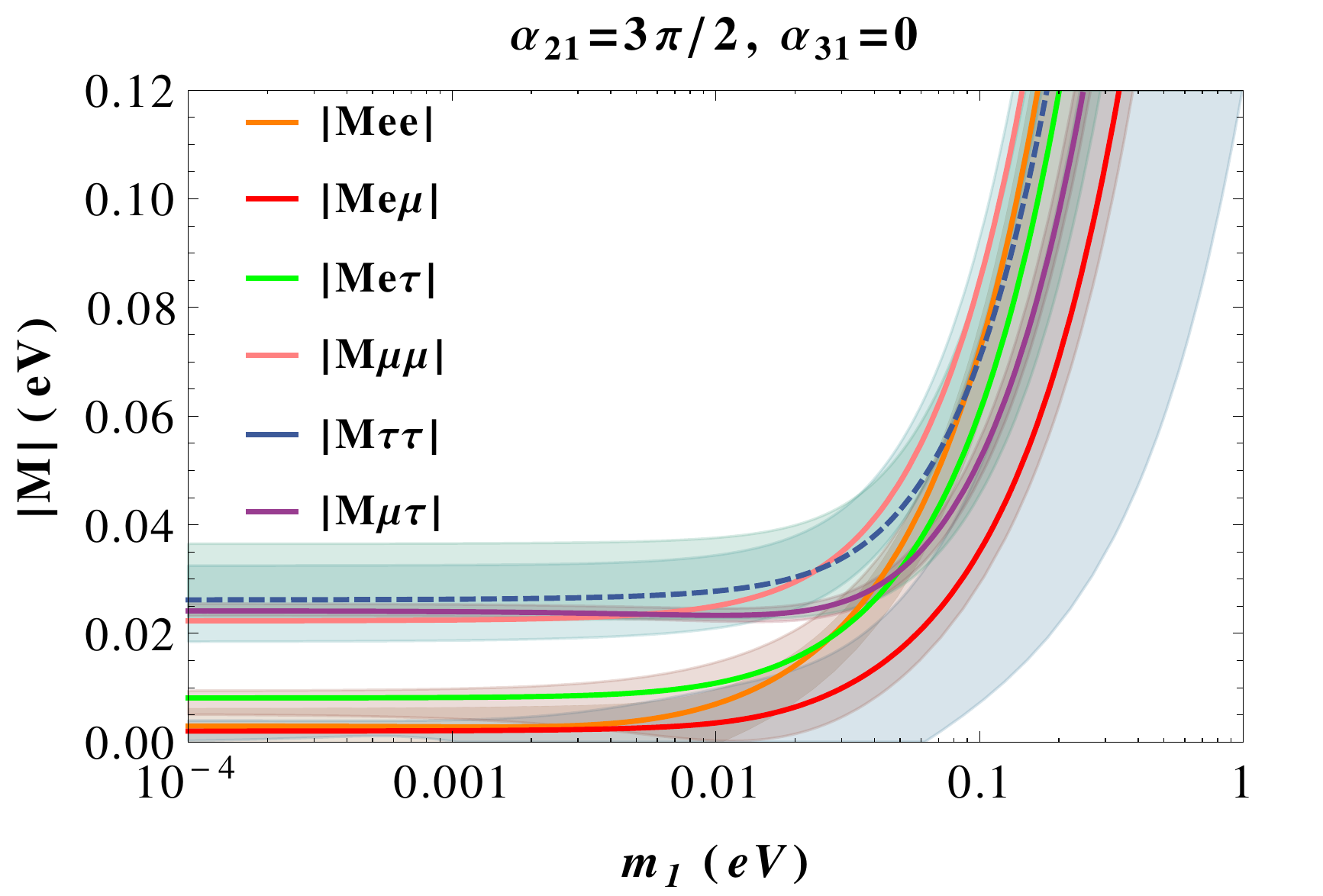}
   \includegraphics[width=.45\textwidth]{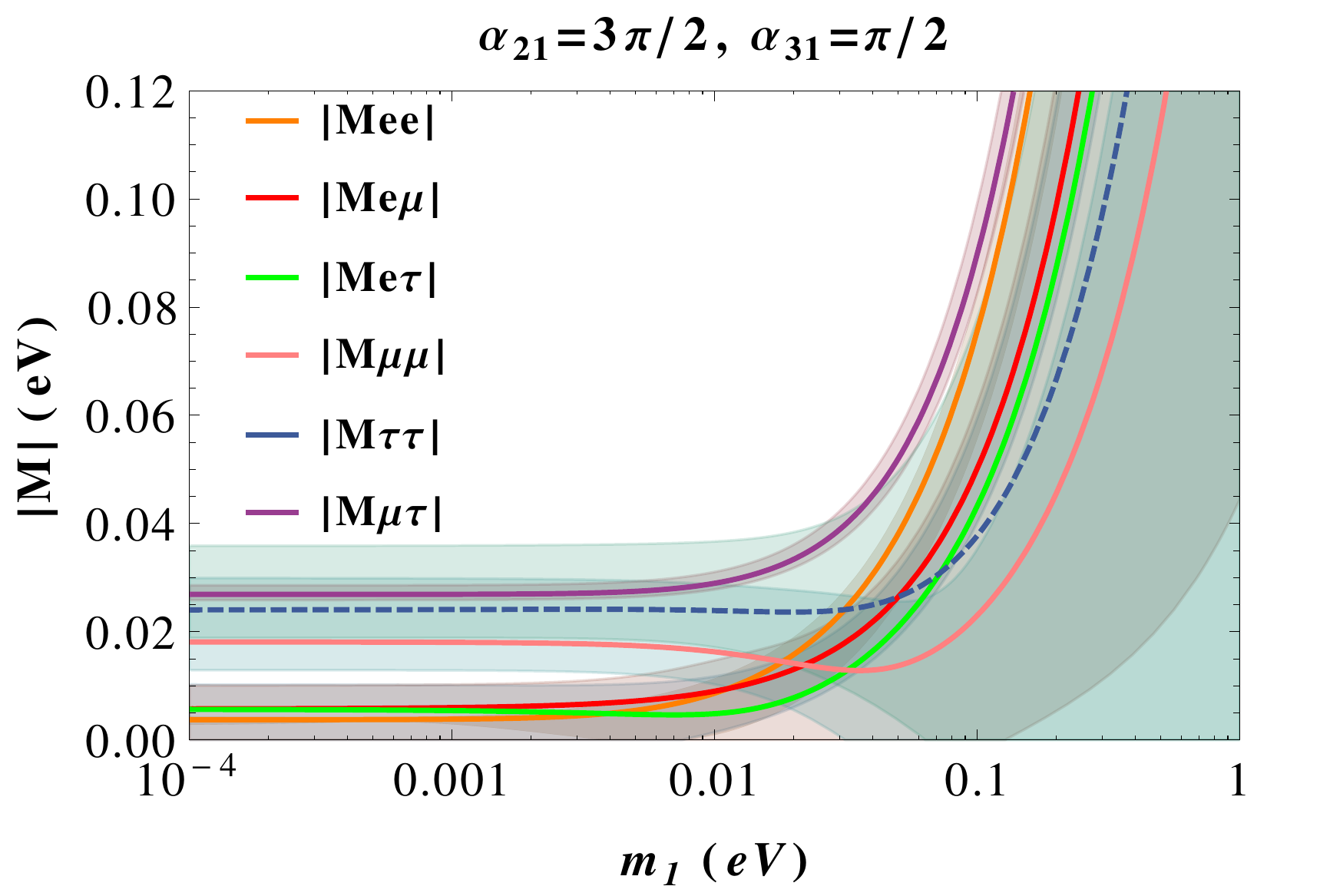}\\
   \includegraphics[width=.45\textwidth]{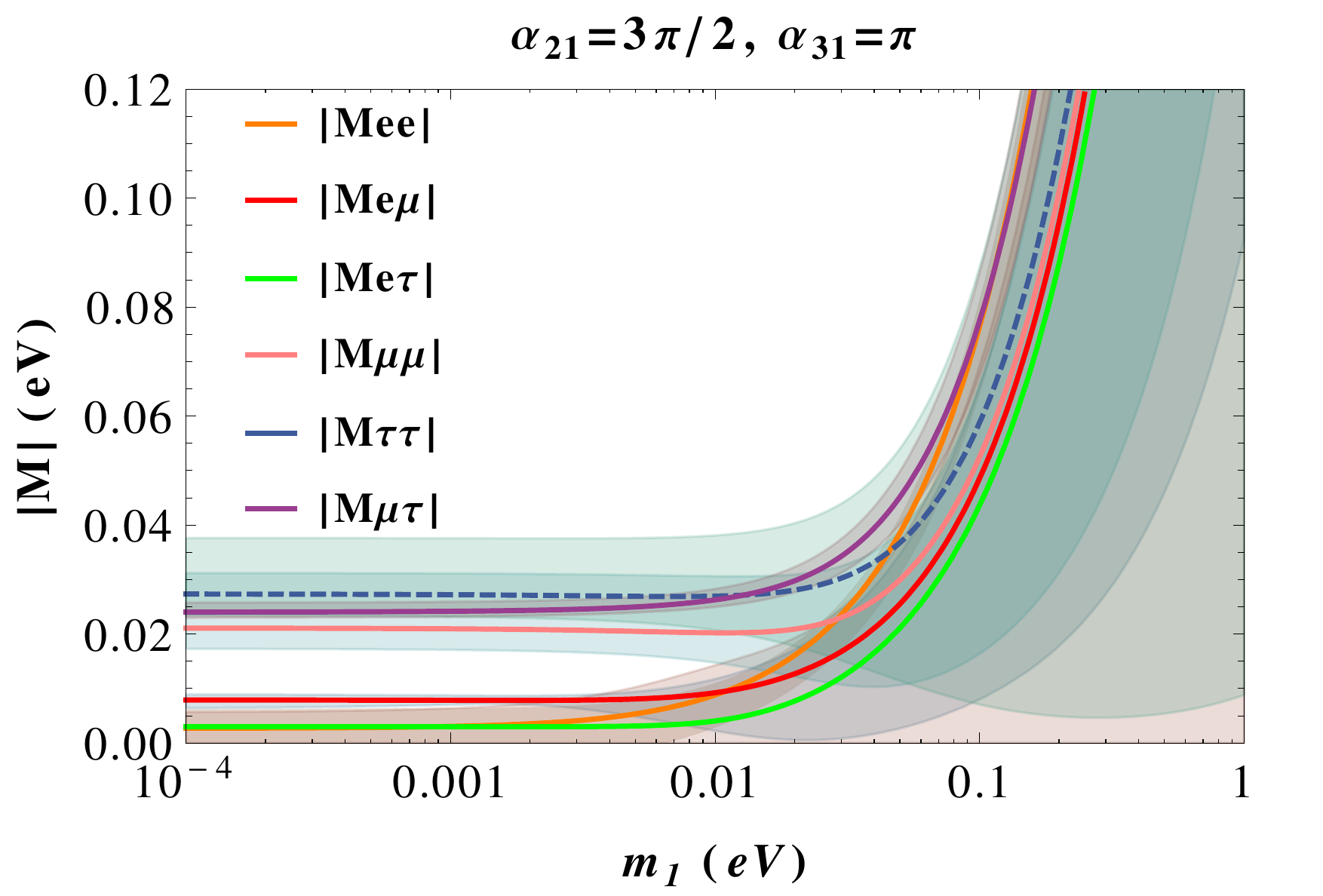}
   \includegraphics[width=.45\textwidth]{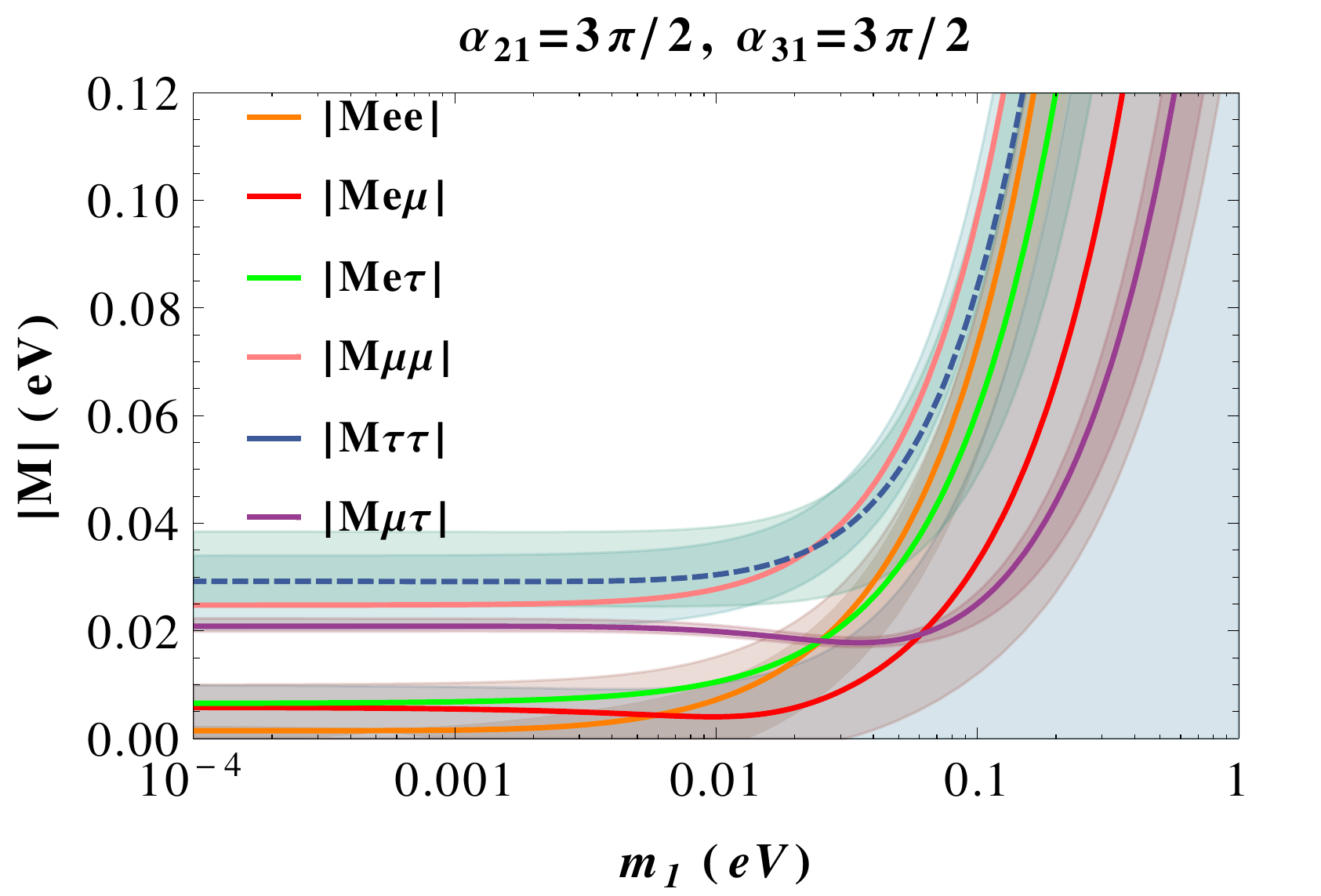}
     \caption{$|\rm M_{\alpha\beta}|$ as a function of the lightest neutrino mass with $(\alpha_{21}, \alpha_{31})$ values equal to $(\pi,0)$, $(\pi,\frac{\pi}{2})$, $(\pi,\pi)$, $(\pi,\frac{3\pi}{2})$, $(\frac{3\pi}{2},0)$, $(\frac{3\pi}{2},\frac{\pi}{2})$, $(\frac{3\pi}{2},\pi)$ , $(\frac{3\pi}{2},\frac{3\pi}{2})$ in the case of the normal ordering.}
  \label{fig:alpha-no2}
\end{minipage}
\end{figure*}

We see that in the cases of non-zero Majorana phases, the general division into three regions according to the stability of the relative magnitudes of different $|\rm M_{\alpha\beta}|$ is applicable. However, the other two major features in the zero-Majorana-phases case, that is, the magnitudes grouping and the $\mu$-$\tau$ symmetry, are changed.

The grouping effect is barely preserved in Region \uppercase\expandafter{\romannumeral1}. In Region \uppercase\expandafter{\romannumeral3}, only the four cases with zero $\alpha_{21}$ exhibit a grouping effect. The $\mu$-$\tau$ symmetric relations, that is, $|\rm M_{e\mu}|\simeq|\rm M_{e\tau}|,\quad |\rm M_{\mu\mu}|\simeq|\rm M_{\tau\tau}|$ work approximately in Region \uppercase\expandafter{\romannumeral1}, and in the cases with $\alpha_{21}=0$ in Region \uppercase\expandafter{\romannumeral3}. It is also interesting to note that $|\rm M_{\mu\tau}|$ grows with $m_1$ more evidently when the Majorana phases are non-zero. Additionally, the error ranges are in general much larger in Region \uppercase\expandafter{\romannumeral3}, due to the increase in complexity when the Majorana phases are non-zero.

\section{Conclusion}
\label{sec:conclusion}
We study the implications for the Majorana neutrino mass matrix given by the current data systematically. The whole set of the oscillation data given by the global fit~\cite{glb_fit,fogli2013} is used to determine the structure of the Majorana neutrino mass matrix up to the lightest neutrino mass.

Since the same subject has been investigated several times before~\cite{mm1,mm2,Grimus:2012ii,Bertuzzo:2013ew}, we would like to emphasize the necessity of the investigation here. Although these works all intend to reconstruct the Majorana neutrino mass matrix, they differ in methodology, so can be used to cross-check the results. We do it in an analytic way. The input here includes all the information given by the global fit result. Especially, we include the value of the Dirac CP-violating phase $\delta$, while the other works leave it unconstrained. The merit here is that we get to a circumstance with least unknown parameters. The demerit is that this treatment relies heavily on the global fit results. To justify this treatment, we use two sets of  global fit results to make comparison, and we find the results are consistent. The only assumption we made is to set the two unknown Majorana phases to zero. We also investigate the non-trivial Majorana phases as a supplement.

Our primary intention is to see the constraints on the structure of the Majorana mass matrix given by the current data, while the other works emphasize the correlations on the matrix elements. By exhibiting the $|\rm M_{\alpha\beta}|$ in the same plot, one can read the dominant structure at given $m_{\rm{min}}$ and make a quick comparison with the model result. Possible texture zeros can be observed directly, too. It is convenient to see the relative magnitudes of different $|\rm M_{\alpha\beta}|$, which can also be used to check the correlations found by the other works.

We make a simple division to the range of $m_{\rm{min}}$ by recognizing the differences in stability of the relative magnitudes of $|\rm M_{\alpha\beta}|$. Then, we discuss the regional characteristics.

We observe a grouping effect in Region \uppercase\expandafter{\romannumeral1} and Region \uppercase\expandafter{\romannumeral3} and an approximating $\mu$-$\tau$ exchange symmetry in all the three regions of both the normal and the inverted orderings.

Some examples of simple parameterizations are listed to exhibit the structures evidently. These parameterizations can be viewed as starting points for a bottom-up way of model building.

We present the results with clear bounds on each matrix element $|\rm M_{\alpha\beta}|$. By taking the bounds seriously, especially the non-trivial lower bounds, we propose six types of texture non-zeros with one-to-one correspondence to each region in the two ordering. These texture non-zeros can be used to make comparison with the texture-zero models.

In extreme cases, the dominant structure is fixed and is shown in Section~\ref{sec:reconstruct}. For non-oscillation data, we find that the cosmology result puts a more stringent constraint than the $0\nu\beta\beta$ result.

It is also observed that the non-zero Majorana phases may change the result in a non-negligible way. It requires precision improvements on these parameters to finally unveil the dominant structure of the Majorana neutrino mass matrix.

\begin{acknowledgements}
We thank Lijing Shao for useful discussions on handling asymmetric errors. This work is supported by National Natural Science Foundation of China (Grant Nos.~11035003 and 11120101004).
\end{acknowledgements}

\appendix

\section{The expressions of the entries of the Majorana neutrino mass matrix}\label{sec:elements}
\begin{eqnarray}
 \mathrm{M}_{ee}&=&c_{13}^2 c_{12}^2 m_1+c_{13}^2 s_{12}^2 m_2 e^{-i \alpha_{21}}+s_{13}^2 m_3 e^{i(2\delta-\alpha_{31})};\nonumber\\
 \mathrm{M}_{e\mu}&=&c_{13} c_{12}(-c_{23}s_{12}-s_{23}c_{12}s_{13}e^{-i\delta})m_1\nonumber\\
         &+&c_{13} s_{12}(c_{23}c_{12}-s_{23}s_{12}s_{13}e^{-i\delta})m_2e^{-i\alpha_{21}}\nonumber\\
         &+&c_{13}s_{23}s_{13}m_3e^{i(\delta-\alpha_{31})};\nonumber\\
 \mathrm{M}_{e\tau}&=&c_{13} c_{12}(s_{23}s_{12}-c_{23}c_{12}s_{13}e^{-i\delta})m_1\nonumber\\
         &+&c_{13} s_{12}(-s_{23}c_{12}-c_{23}s_{12}s_{13}e^{-i\delta})m_2e^{-i\alpha_{21}}\nonumber\\
         &+&c_{13}c_{23}s_{13}m_3e^{i(\delta-\alpha_{31})};\nonumber\\
 \mathrm{M}_{\mu\mu}&=&(-c_{23}s_{12}-s_{23}c_{12}s_{13}e^{-i\delta})^2m_1\nonumber\\
         &+&(c_{23}c_{12}-s_{23}s_{12}s_{13}e^{-i\delta})^2m_2e^{-i\alpha_{21}}\nonumber\\
         &+&c_{13}^2s_{23}^2m_3e^{-i\alpha_{31}};\nonumber\\
 \mathrm{M}_{\mu\tau}&=&(s_{23}s_{12}-c_{23}c_{12}s_{13}e^{-i\delta})(-c_{23}s_{12}-s_{23}c_{12}s_{13}e^{-i\delta})m_1\nonumber\\
          &+&(-s_{23}c_{12}-c_{23}s_{12}s_{13}e^{-i\delta})(c_{23}c_{12}-s_{23}s_{12}s_{13}e^{-i\delta})m_2e^{-i\alpha_{21}}\nonumber\\
          &+&c_{13}^2c_{23}s_{23}m_3e^{-i\alpha_{31}};\nonumber\\
 \mathrm{M}_{\tau\tau}&=&(s_{23}s_{12}-c_{23}c_{12}s_{13}e^{-i\delta})^2m_1\nonumber\\
         &+&(-s_{23}c_{12}-c_{23}s_{12}s_{13}e^{-i\delta})^2m_2e^{-i\alpha_{21}}\nonumber\\
         &+&c_{13}^2c_{23}^2m_3e^{-i\alpha_{31}}.\nonumber
\end{eqnarray}

\section{The same procedure with Fogli et al. data}
\label{sec:fogli}

\begin{figure*}
\begin{minipage}{\textwidth}
   \centering
  \includegraphics[width=.45\textwidth]{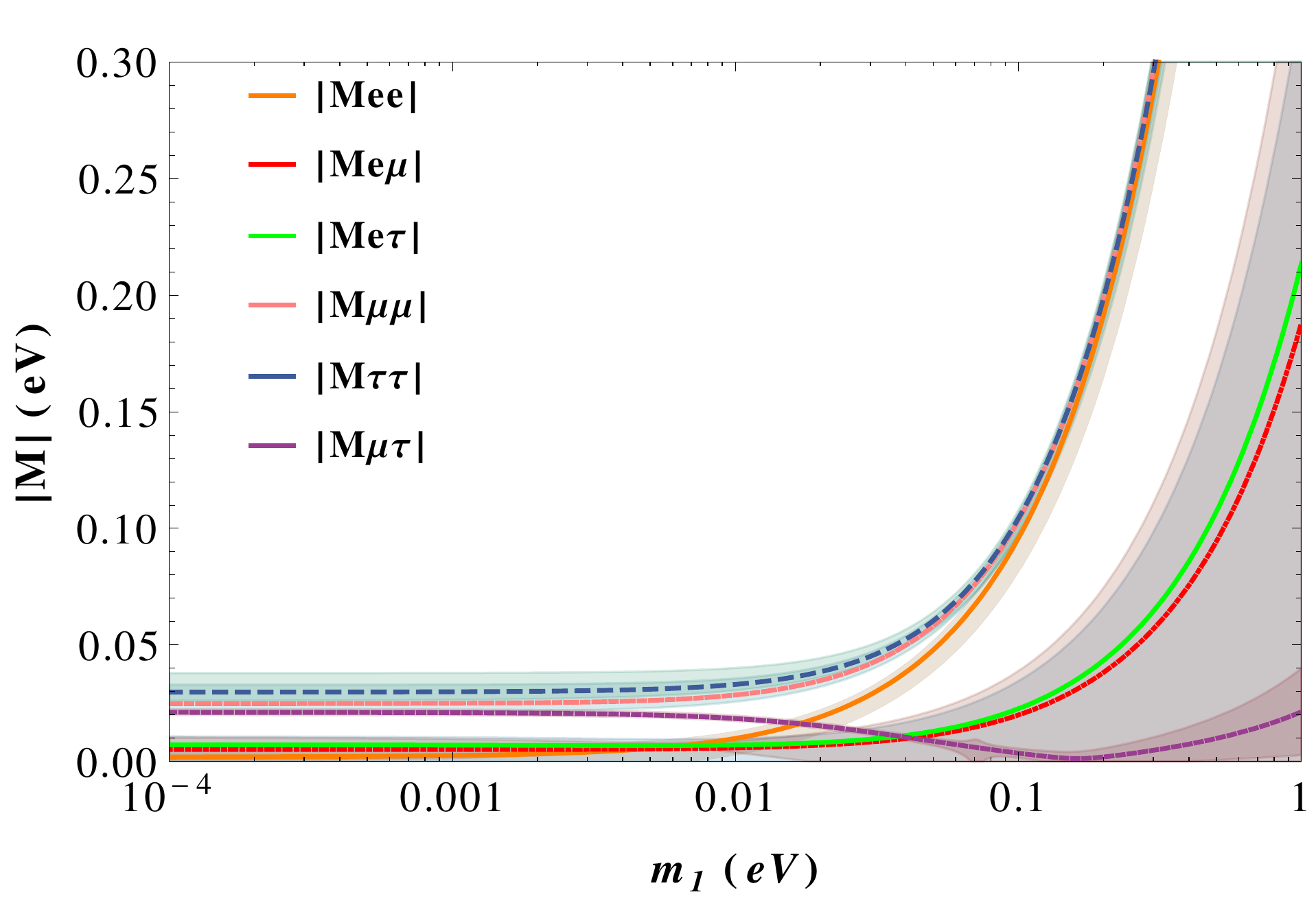}
  \includegraphics[width=.45\textwidth]{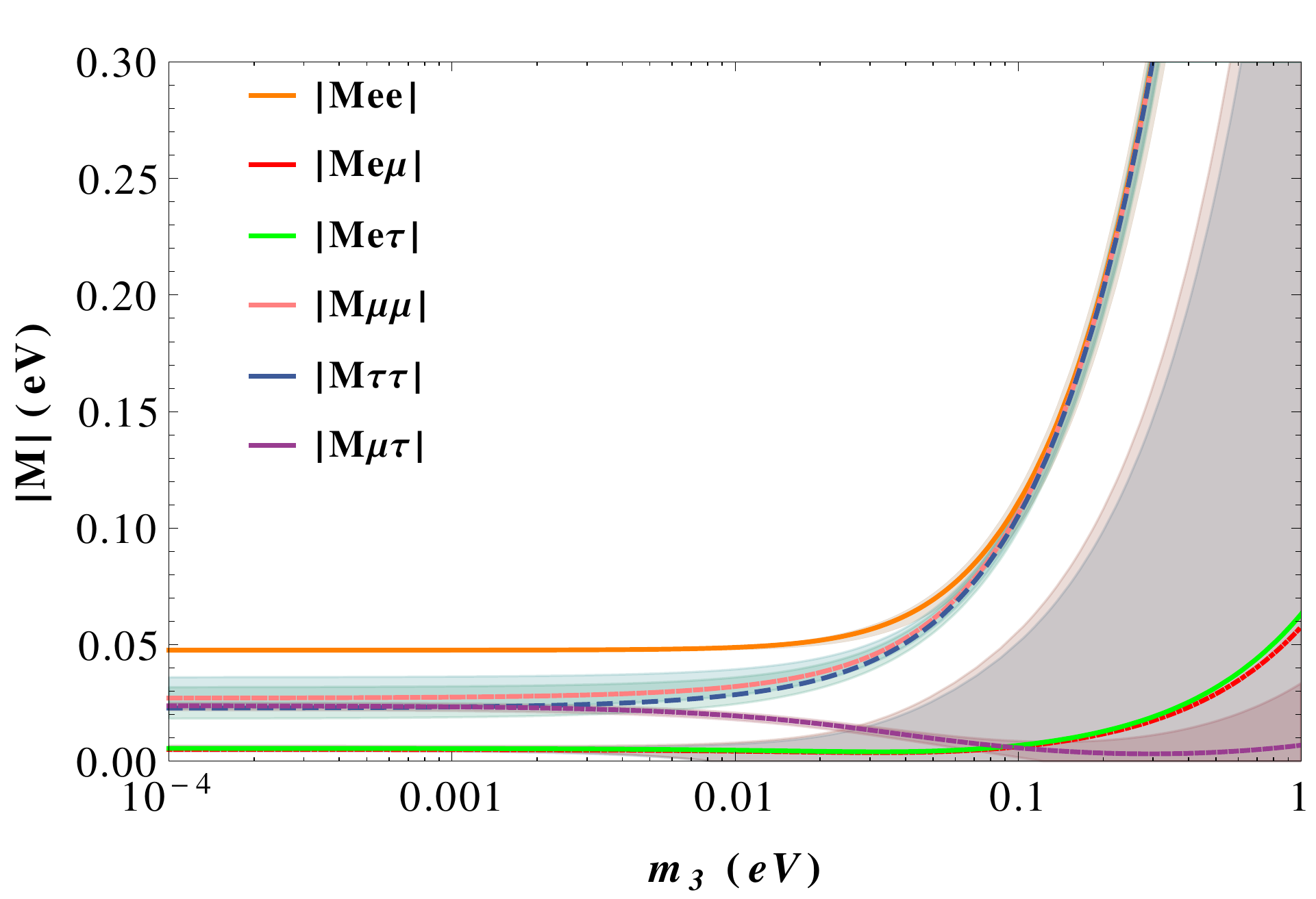}
        \caption{\label{fig:iii} $|\rm M_{\alpha\beta}|$ as a function of the lightest neutrino mass, using ref.~\cite{Fogli:2012ua} as input. The left one corresponds to the normal ordering, while the right one corresponds to the inverted ordering.}
\end{minipage}
\end{figure*}

We perform the same procedure with the global fit result from ref.~\cite{fogli2013}. We list the input in Table~\ref{tab:ii}, where $\Delta m^2=m_3^2-(m_1^2+m_2^2)/2$. The result is shown in Figure~\ref{fig:iii}.

We see that Region \uppercase\expandafter{\romannumeral1} exhibits the same characteristics as the results obtained using the ref.~\cite{glb_fit} as an input, while Region \uppercase\expandafter{\romannumeral3} shows much larger error ranges than in Figure~\ref{fig:i}, which is a result of the mass dependence in $\Delta m^2$ definition.

The relative magnitudes are found to be
\begin{eqnarray}
|\rm M_{\tau\tau}|>|\rm M_{\mu\mu}|>|\rm M_{\mu\tau}|>|\rm M_{e\tau}|>|\rm M_{ee}|>|\rm M_{e\mu}|
\end{eqnarray}
in Region \uppercase\expandafter{\romannumeral1} of the normal ordering. The feature of magnitudes grouping is the same as the results we obtained using ref.~\cite{glb_fit} as an input. The first three are of $\mathcal{O}(10^{-2})\rm~eV$, while the latter three are of $\mathcal{O}(10^{-3})\rm~eV$. $|\rm M_{ee}|$ and $|\rm M_{e\mu}|$ are of different relative magnitudes in comparison with Eq.~\ref{eqn:no}.

The relative magnitudes in the inverted ordering case are
the same as Eq.~\ref{eqn:io}.

The $\mu$-$\tau$ exchange symmetry is also recognized in all the regions of both the orderings.

\section{The effects of the Majorana phases in the case of inverted ordering}
\label{sec:io}

\begin{figure*}
\begin{minipage}{\textwidth}
   \centering
    \includegraphics[width=.45\textwidth]{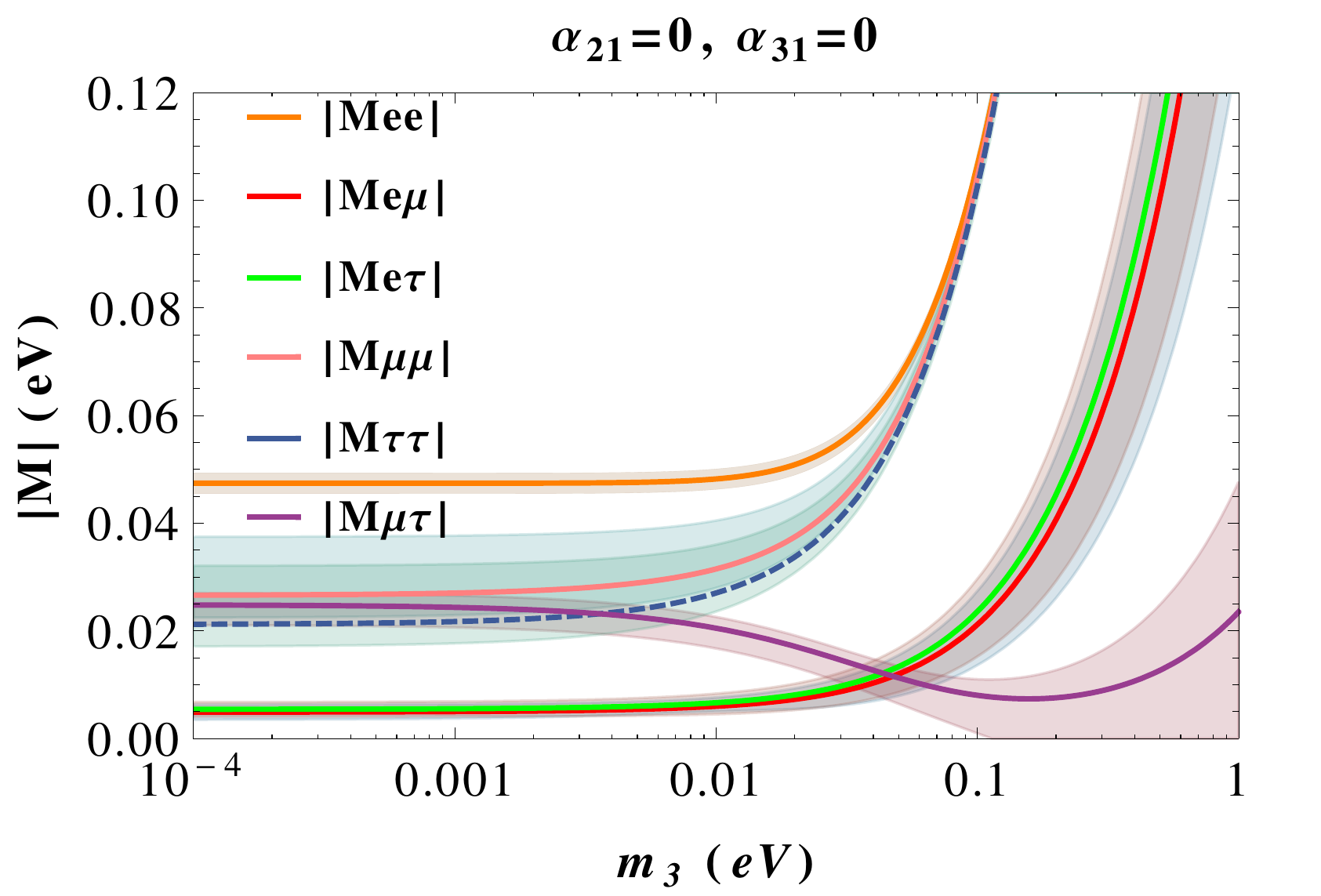}
    \includegraphics[width=.45\textwidth]{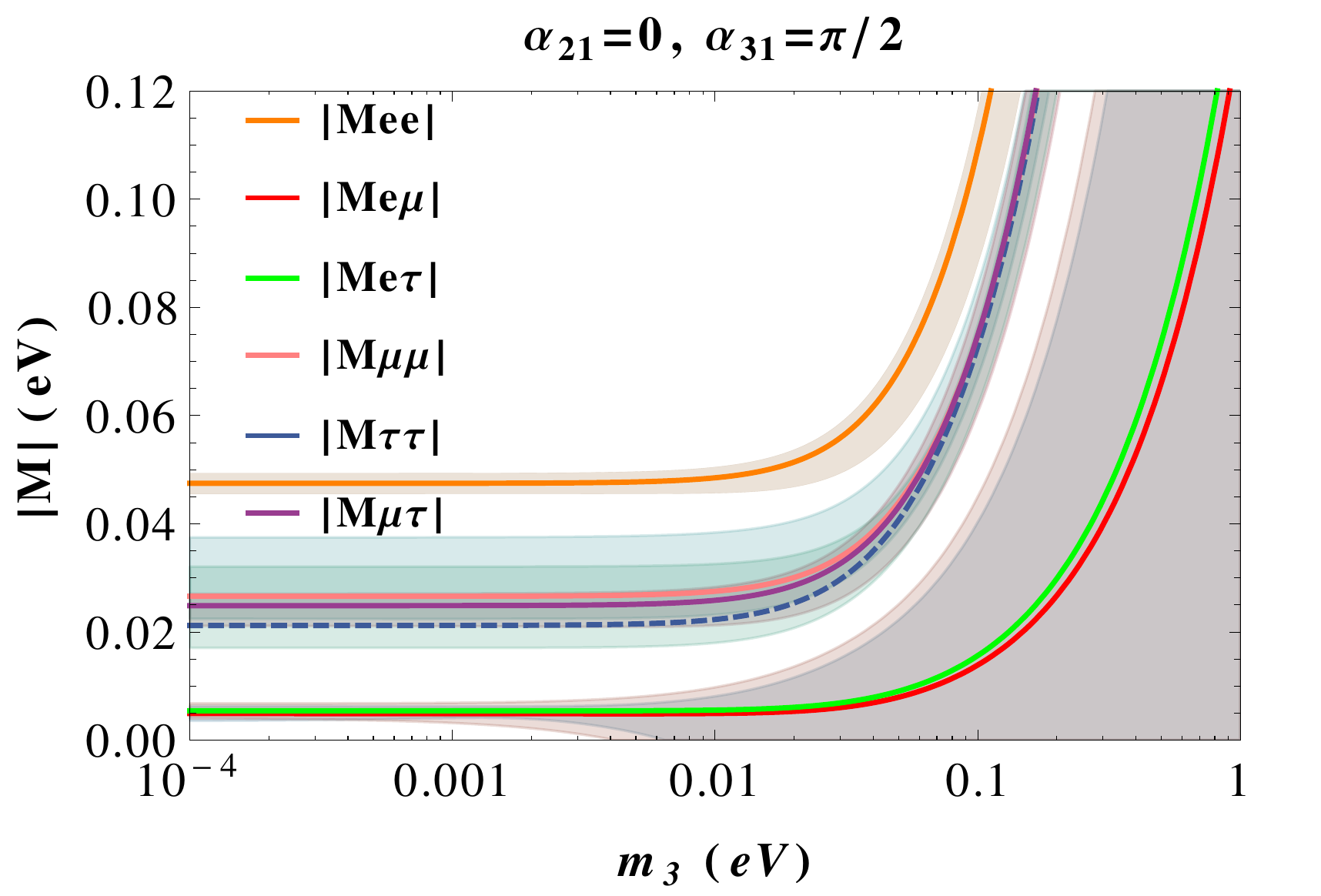}\\
    \includegraphics[width=.45\textwidth]{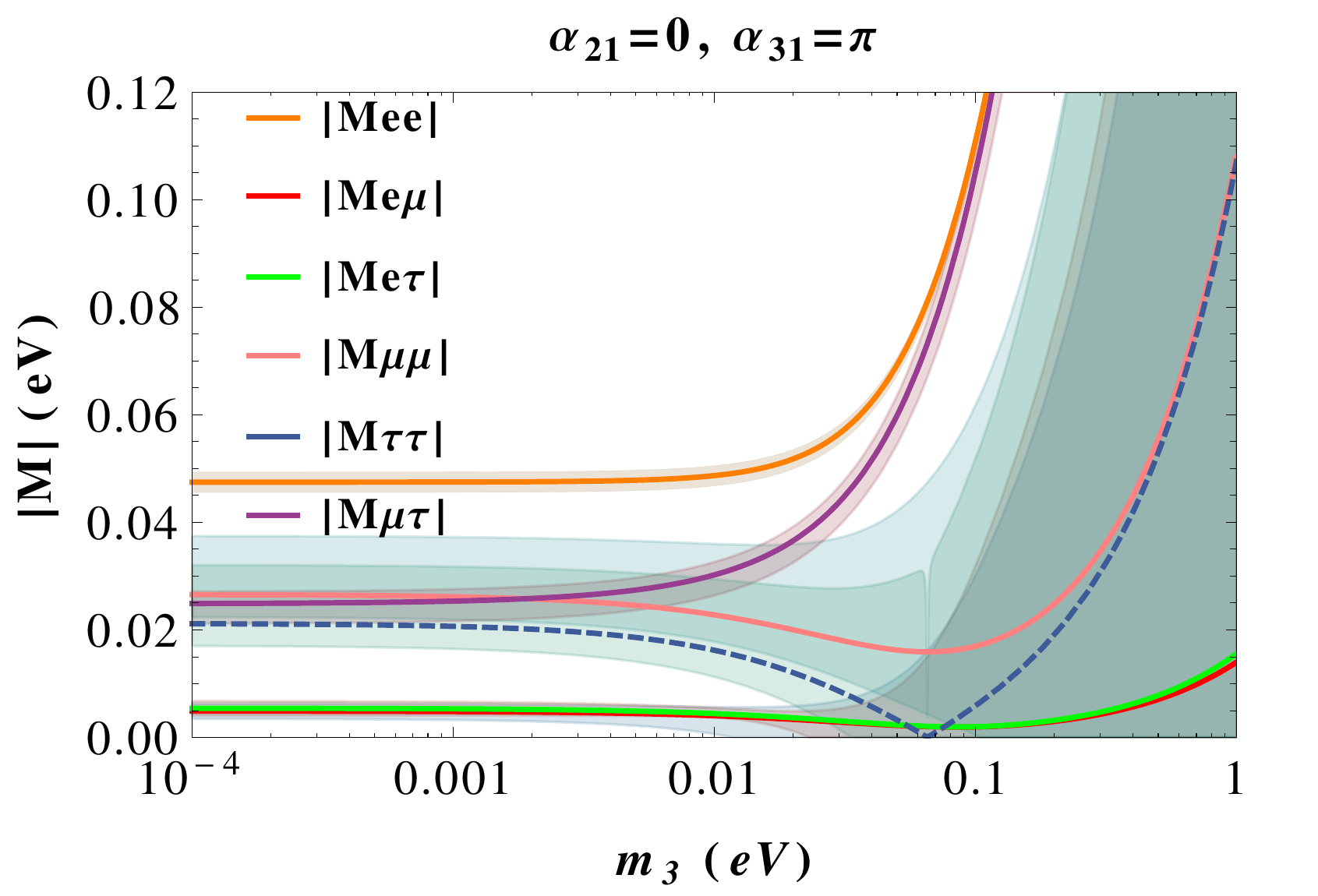}
    \includegraphics[width=.45\textwidth]{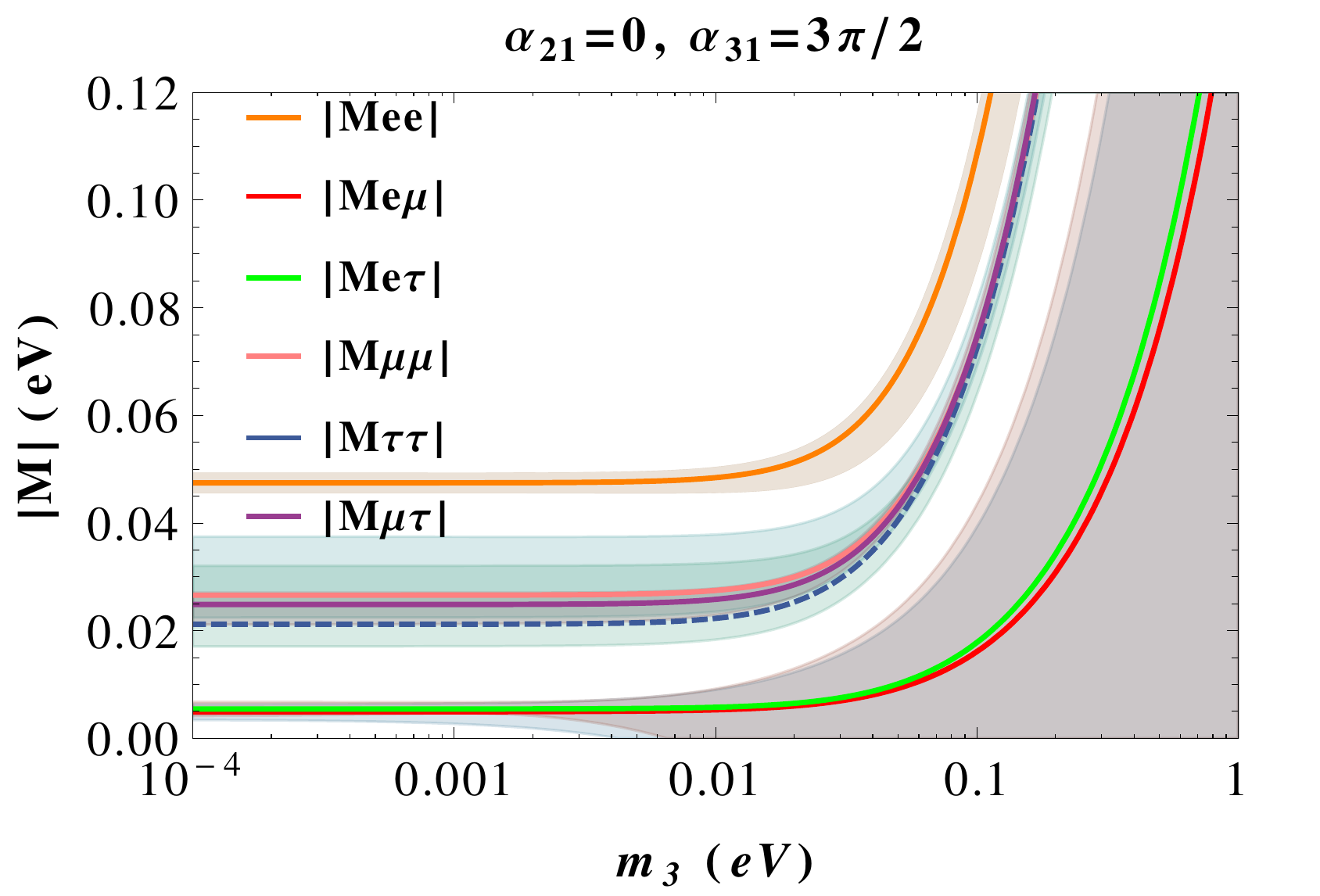}\\
    \includegraphics[width=.45\textwidth]{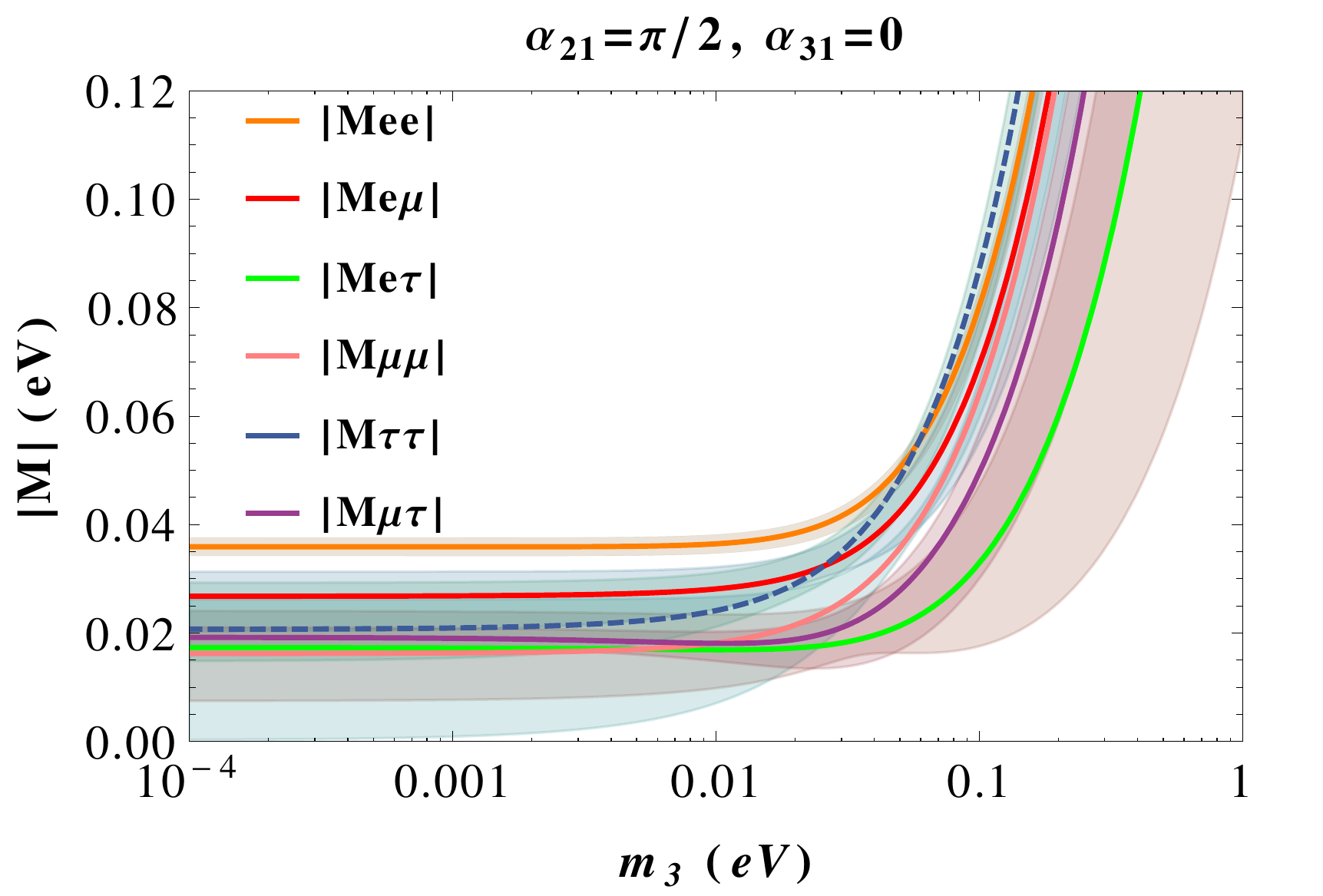}
    \includegraphics[width=.45\textwidth]{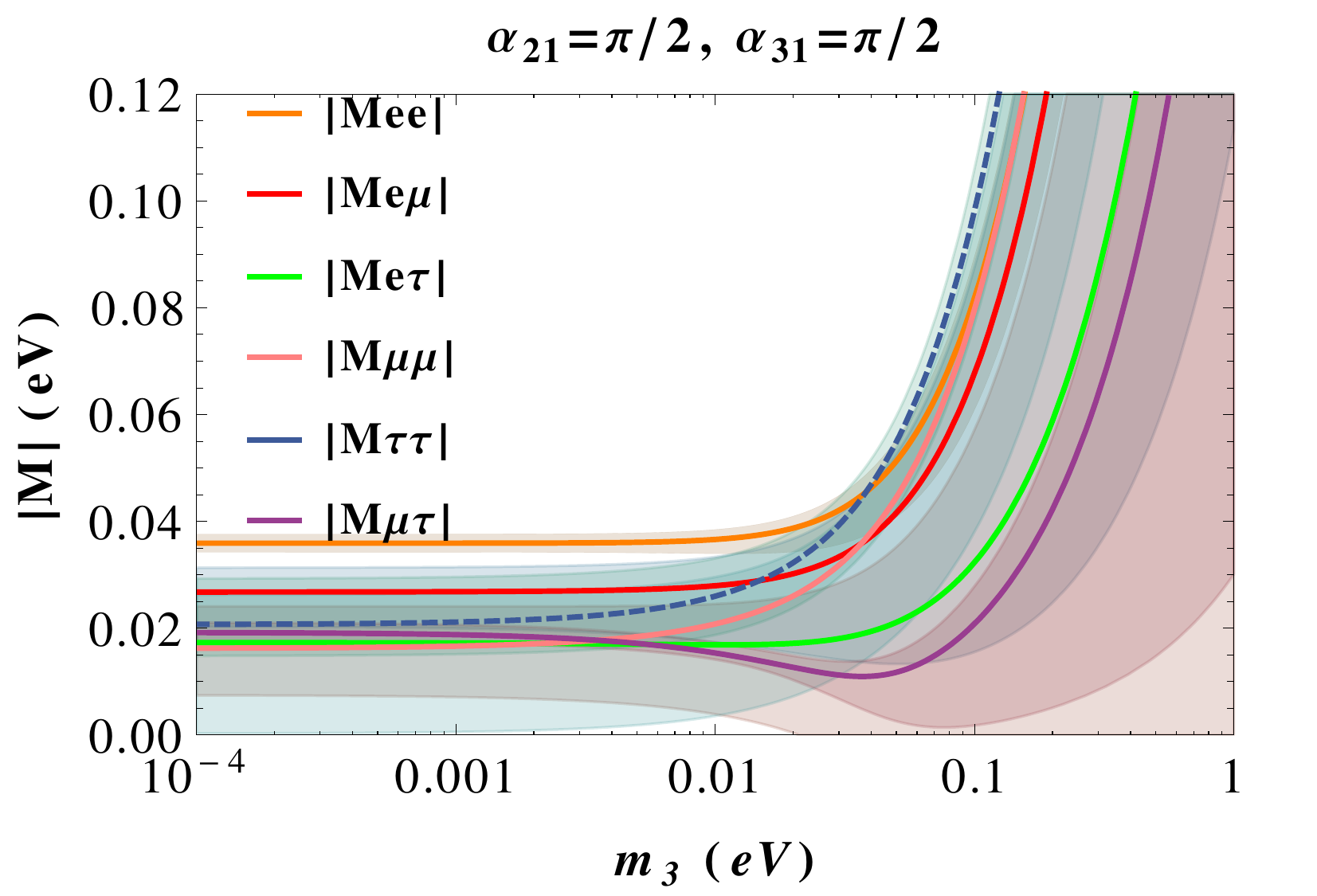}\\
    \includegraphics[width=.45\textwidth]{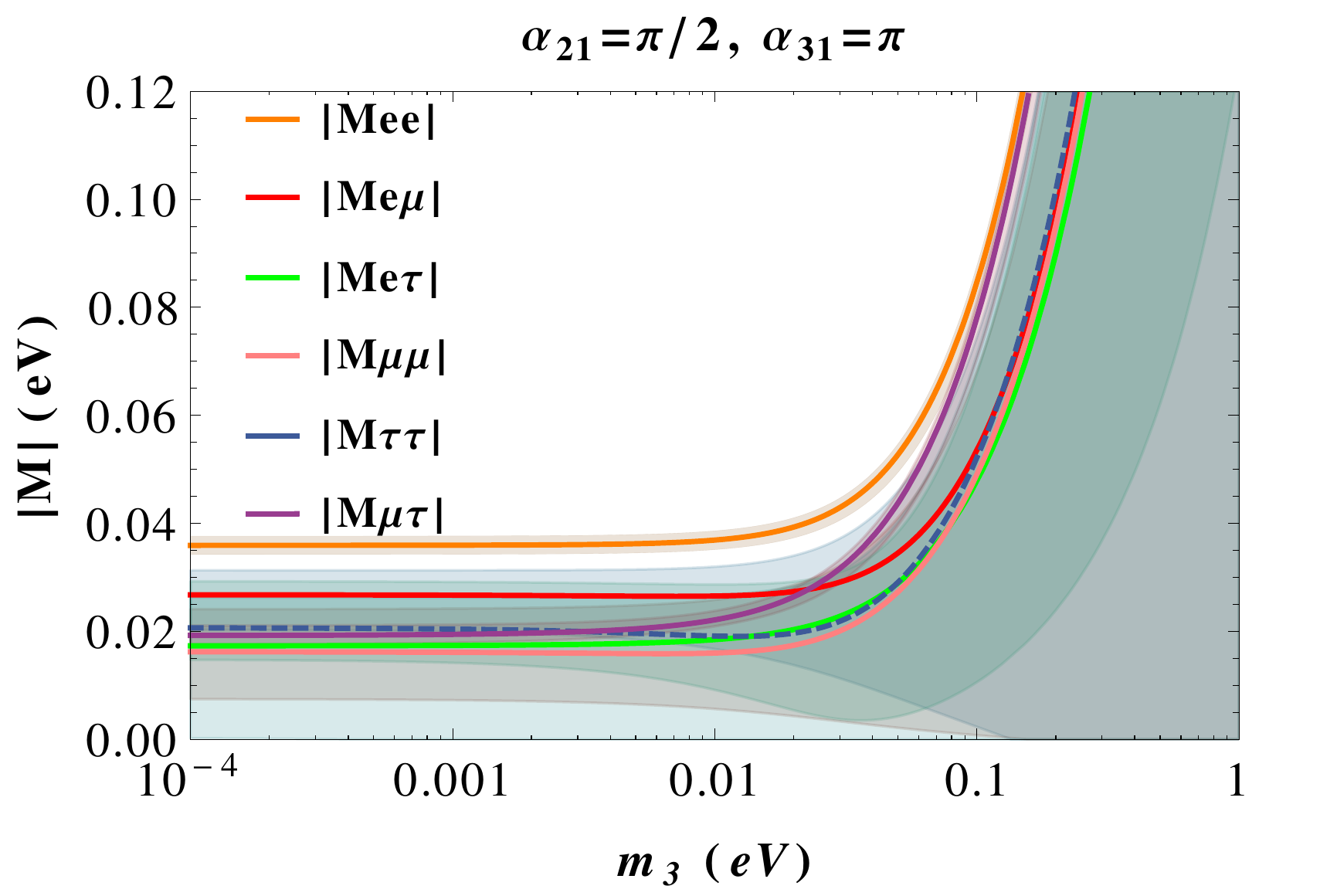}
    \includegraphics[width=.45\textwidth]{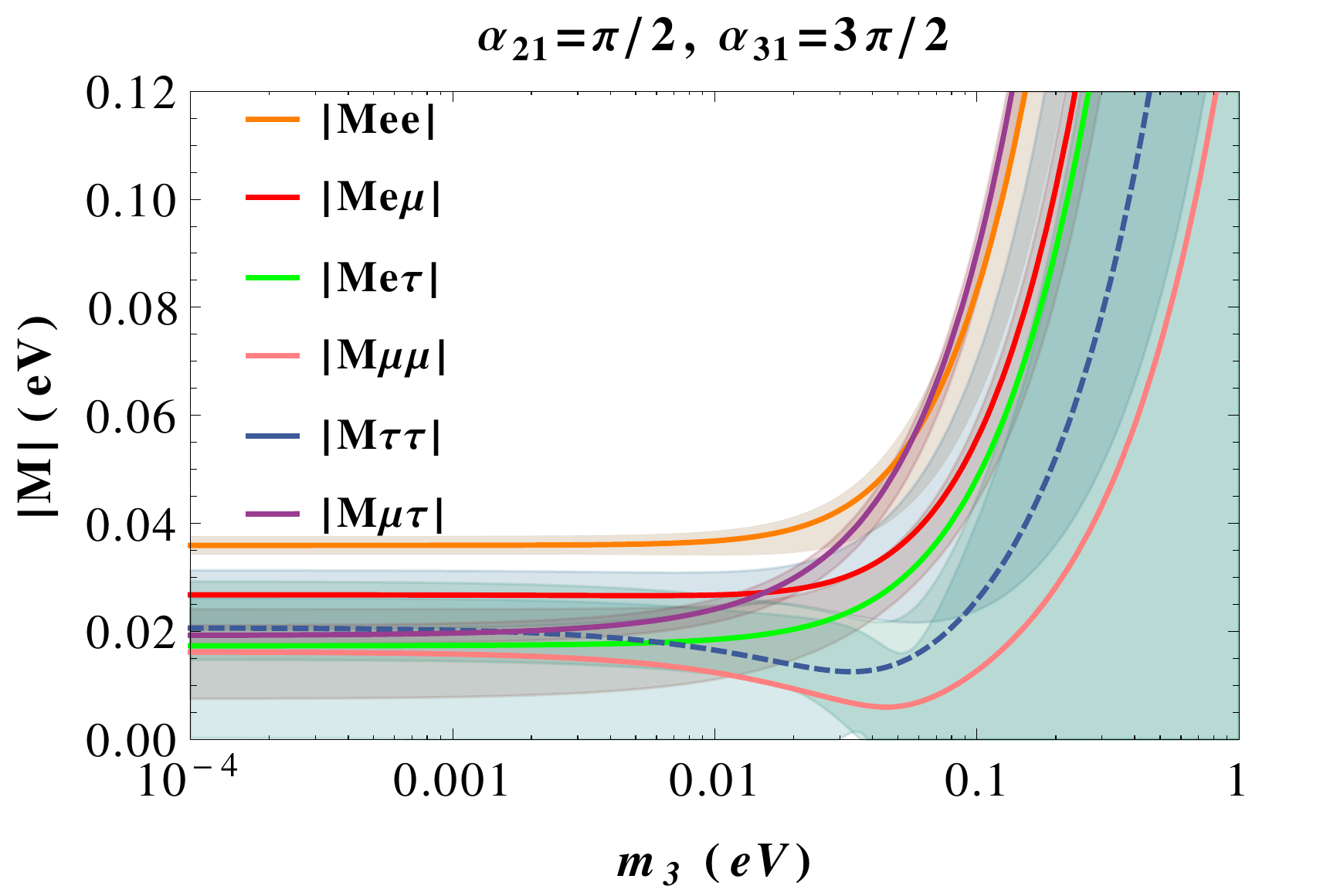}
       \caption{$|\rm M_{\alpha\beta}|$ as a function of the lightest neutrino mass with $(\alpha_{21}, \alpha_{31})$ values equal to  $(0,0)$, $(0,\frac{\pi}{2})$, $(0,\pi)$, $(0,\frac{3\pi}{2})$, $(\frac{\pi}{2},0)$ , $(\frac{\pi}{2},\frac{\pi}{2})$, $(\frac{\pi}{2},\pi)$, $(\frac{\pi}{2},\frac{3\pi}{2})$ in the case of the inverted ordering.}
  \label{fig:io1}
\end{minipage}
\end{figure*}

\begin{figure*}
\begin{minipage}{\textwidth}
   \centering
    \includegraphics[width=.45\textwidth]{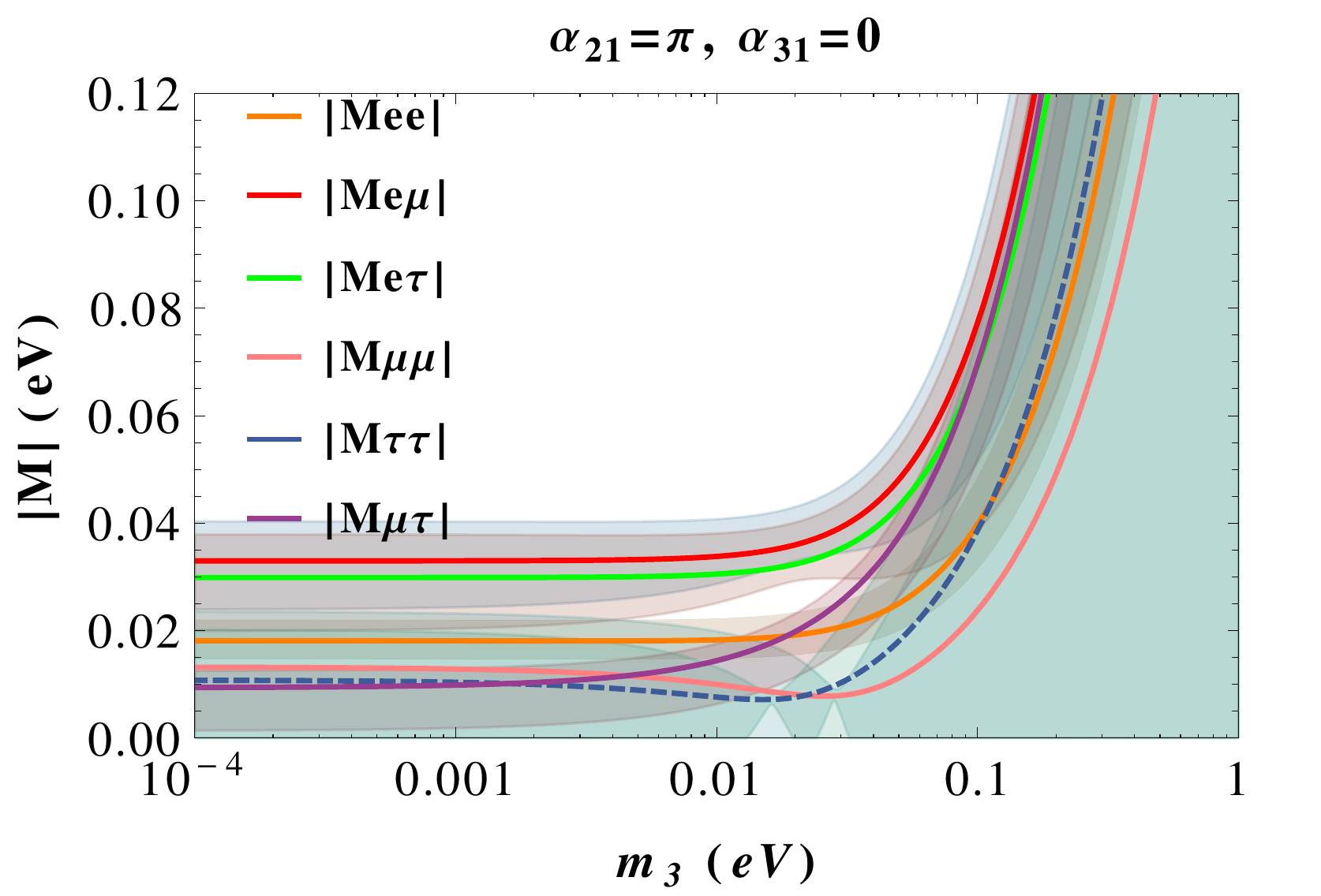}
    \includegraphics[width=.45\textwidth]{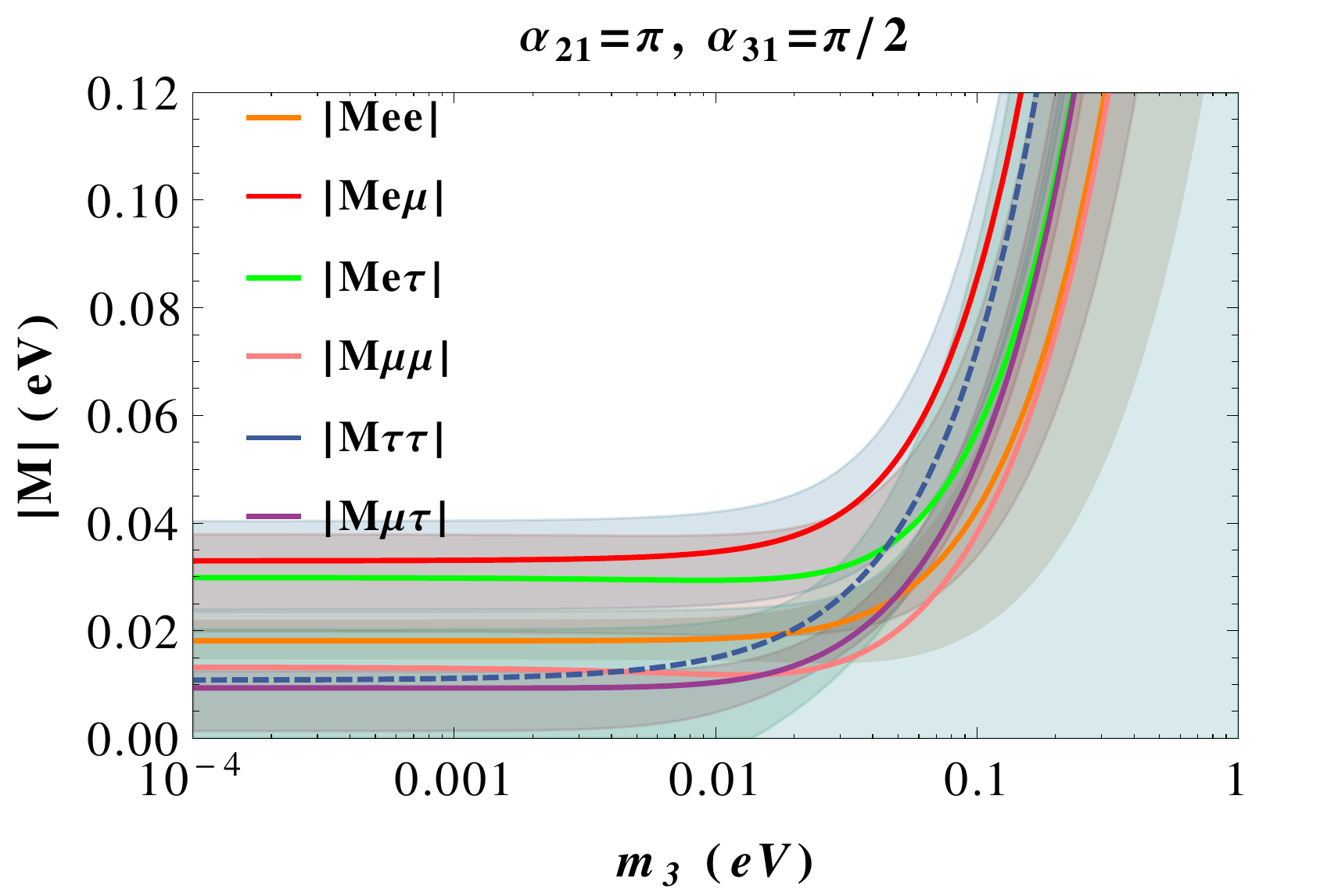}\\
    \includegraphics[width=.45\textwidth]{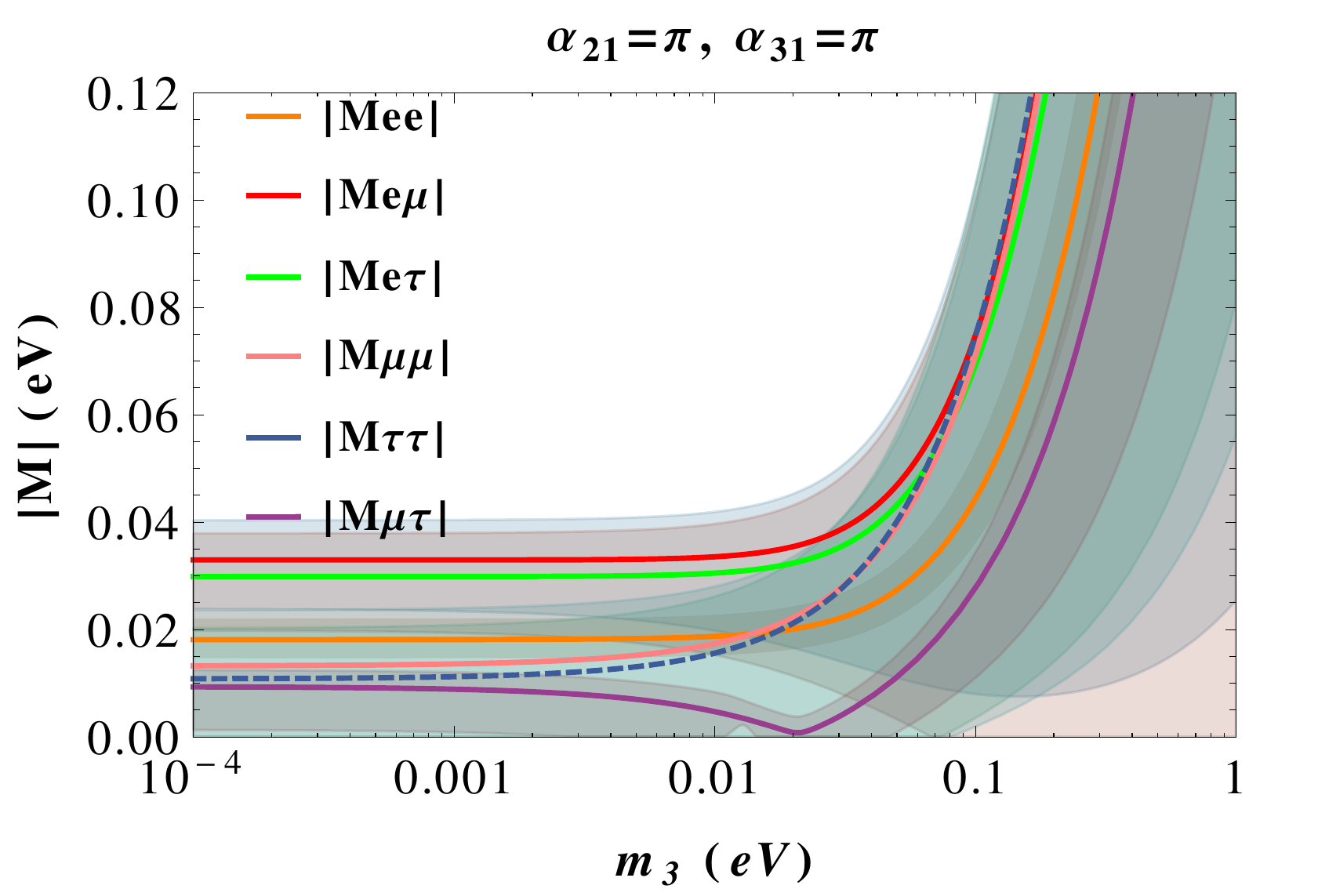}
    \includegraphics[width=.45\textwidth]{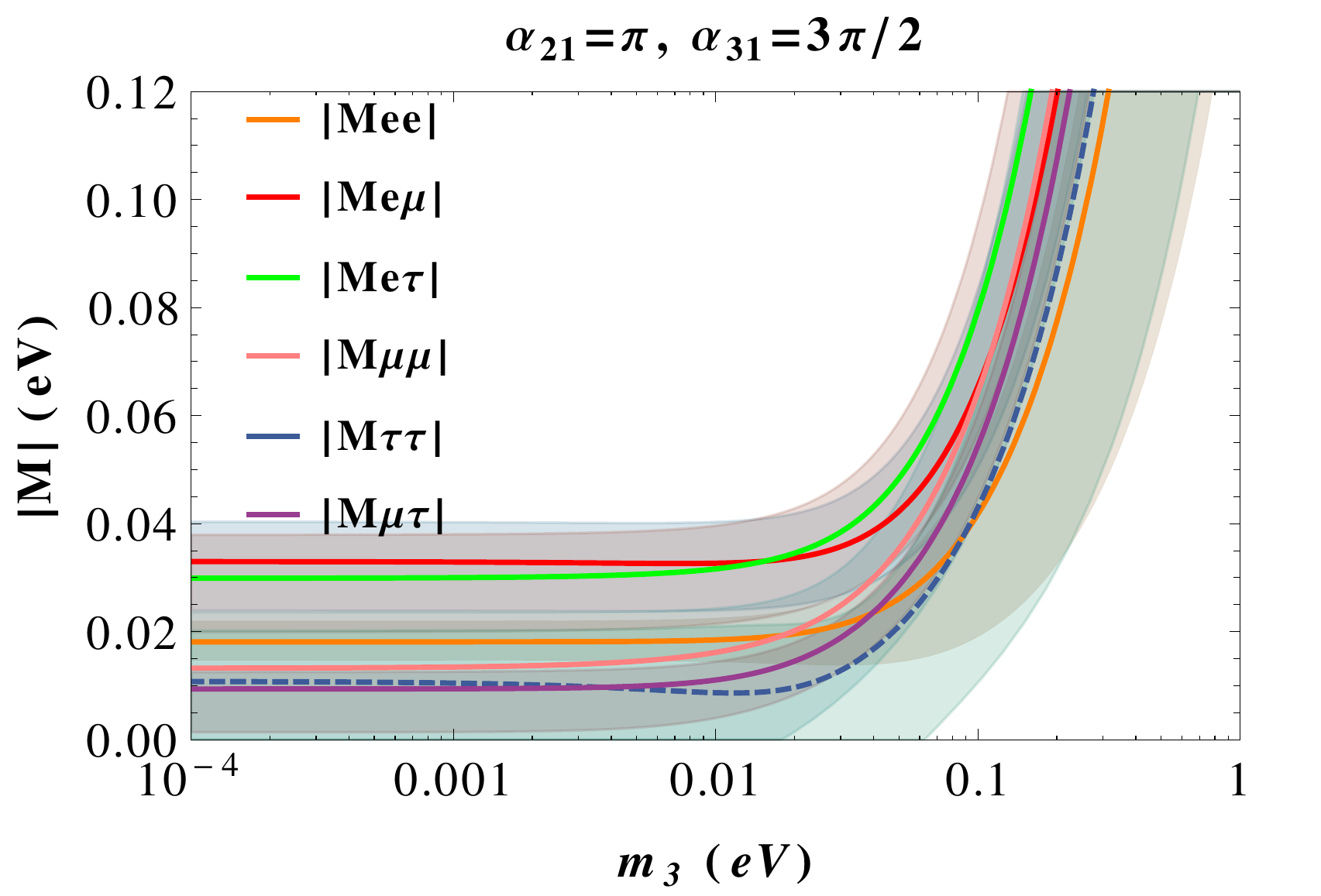}\\
    \includegraphics[width=.45\textwidth]{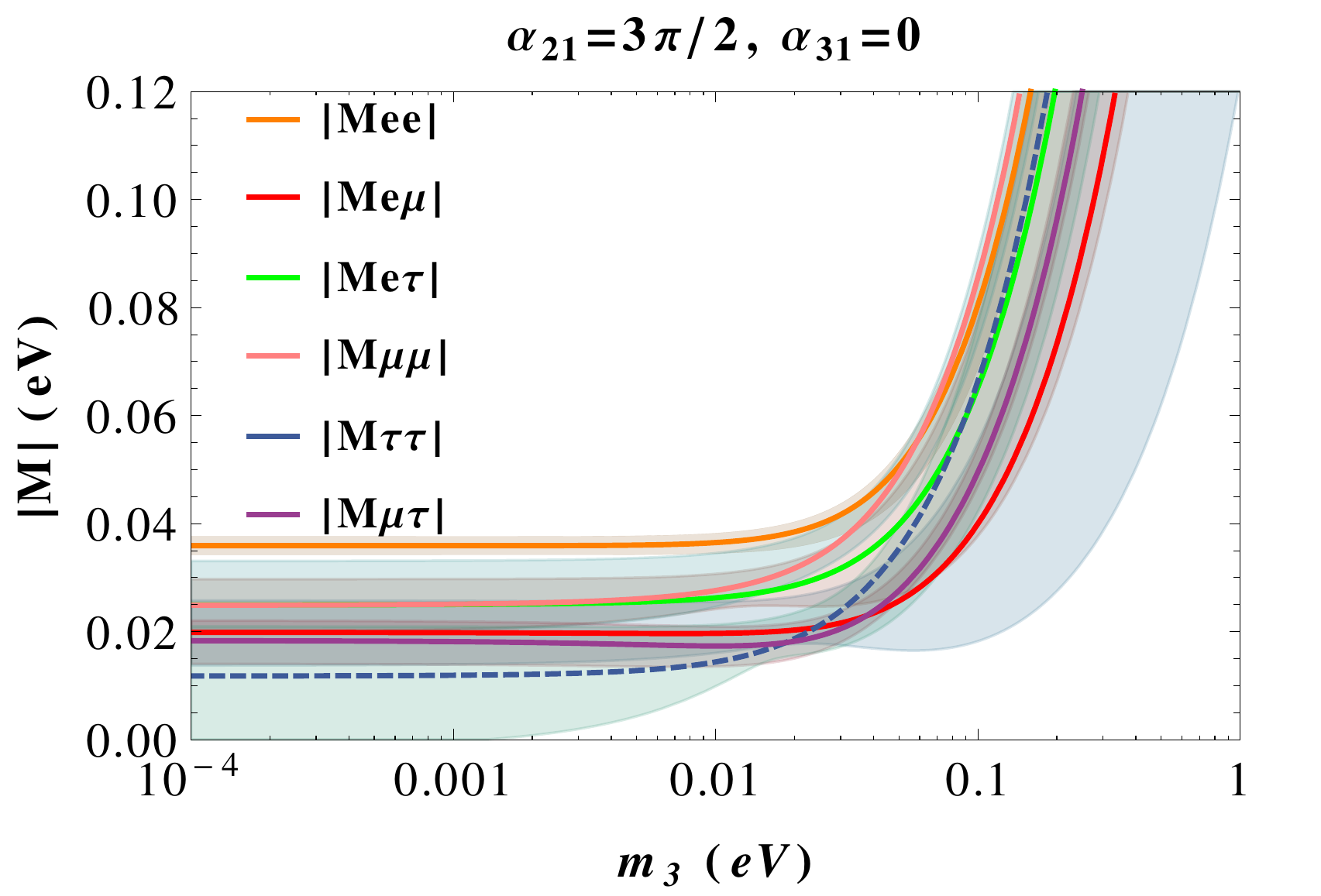}
    \includegraphics[width=.45\textwidth]{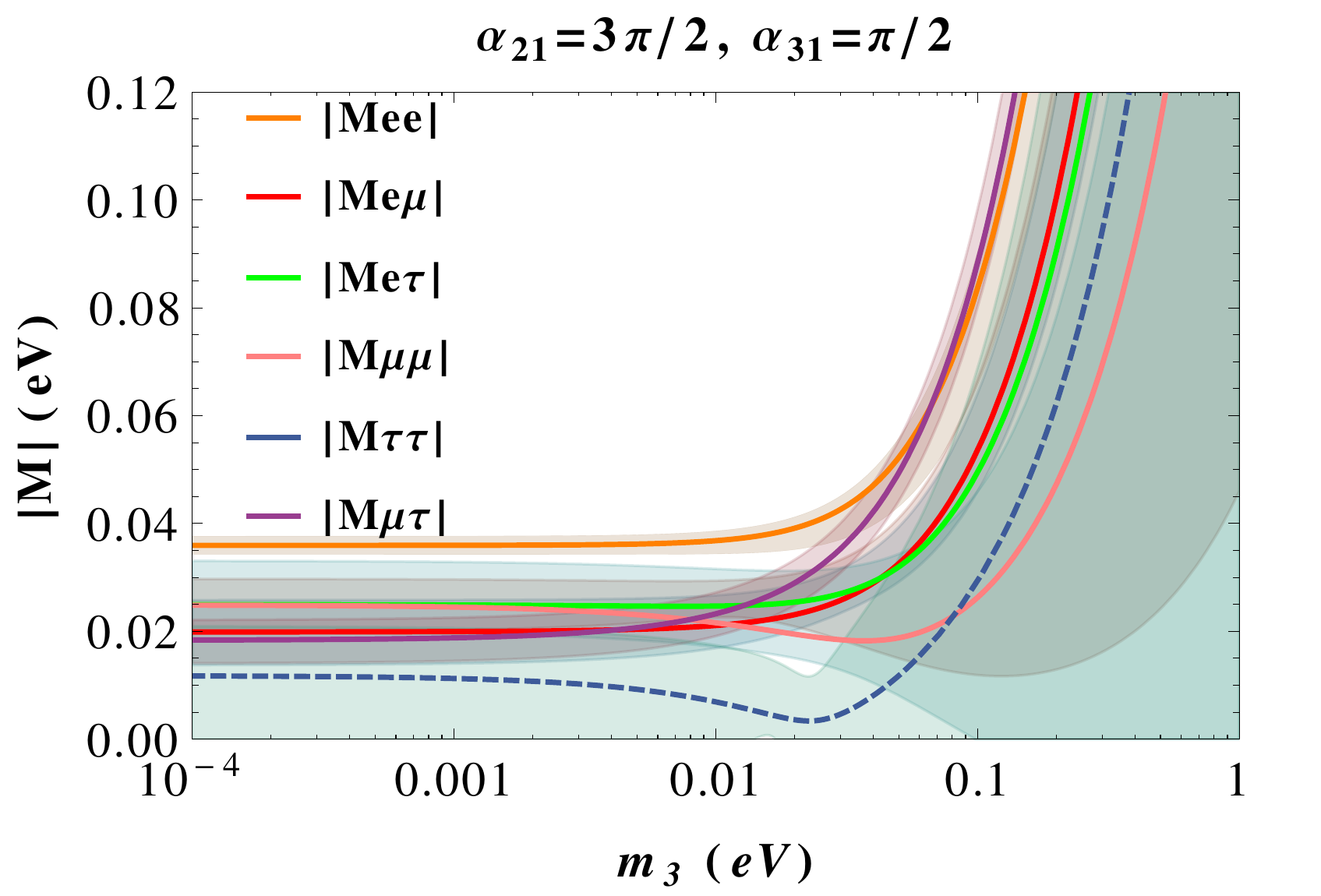}\\
    \includegraphics[width=.45\textwidth]{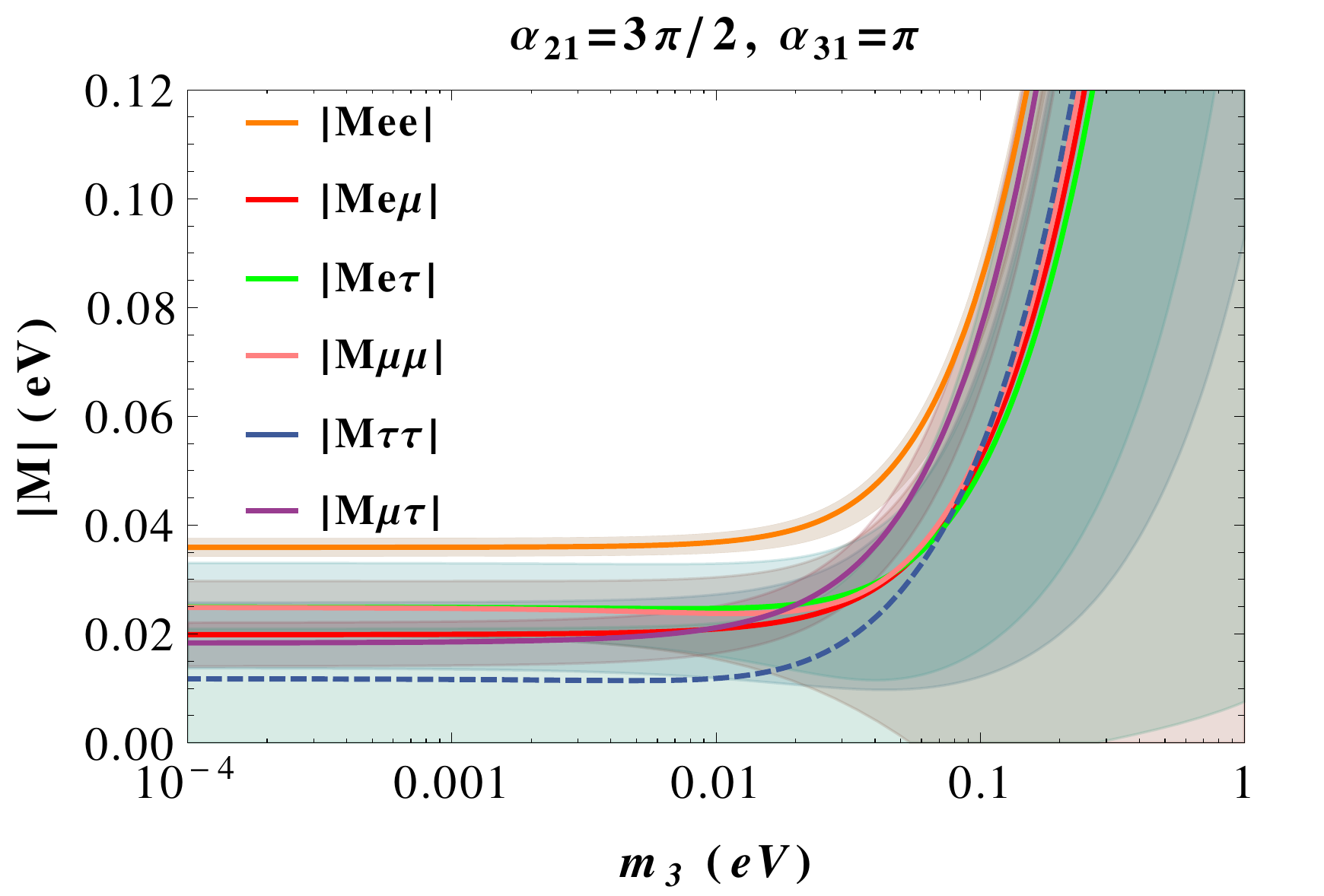}
    \includegraphics[width=.45\textwidth]{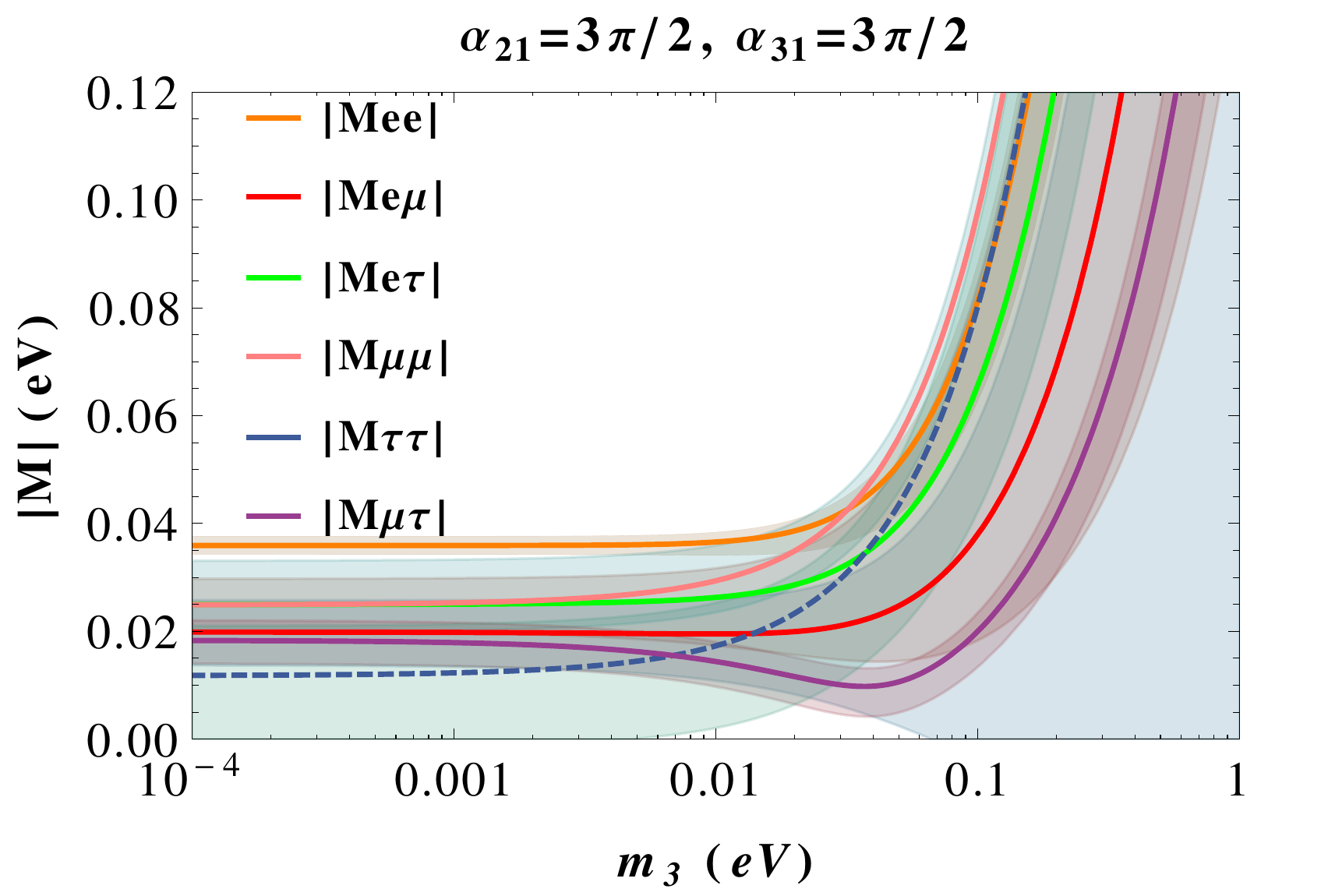}
    \caption{$|\rm M_{\alpha\beta}|$ as a function of the lightest neutrino mass with $(\alpha_{21}, \alpha_{31})$ values equal to $(\pi,0)$, $(\pi,\frac{\pi}{2})$, $(\pi,\pi)$, $(\pi,\frac{3\pi}{2})$, $(\frac{3\pi}{2},0)$, $(\frac{3\pi}{2},\frac{\pi}{2})$, $(\frac{3\pi}{2},\pi)$ , $(\frac{3\pi}{2},\frac{3\pi}{2})$ in the case of the inverted ordering.}
  \label{fig:io2}
\end{minipage}
\end{figure*}

From Figures~\ref{fig:io1}~and~\ref{fig:io2}, we see that the general division into three regions is applicable. The grouping effect is only observed when $\alpha_{21}=0$ in Region \uppercase\expandafter{\romannumeral1} and $(0,\frac{\pi}{2})$ in Region \uppercase\expandafter{\romannumeral3}. The $\mu$-$\tau$ symmetric relations, that is, $|\rm M_{e\mu}|\simeq|\rm M_{e\tau}|,\quad |\rm M_{\mu\mu}|\simeq|\rm M_{\tau\tau}|$ do not serve as good approximations when $\alpha_{21}=0, \frac{3\pi}{2}$. It is interesting to note that when $\alpha_{21}=0, \pi$, that is the CP-conserving values, the $\mu$-$\tau$ symmetric relations work approximately. 
The error ranges are in general large in Region \uppercase\expandafter{\romannumeral3} and $|\rm M_{\mu\tau}|$ grows with $m_3$ evidently.

\end{document}